\documentclass[
  aip,
  author-year,
  reprint,
  onecolumn,
  tightenlines,
  12pt,
  notitlepage,
  amsmath,
  amssymb
]{revtex4-1}
\setcitestyle{square}
\usepackage[pdftex]{graphicx}
\usepackage[pdftex,bookmarks,colorlinks,citecolor=blue,linkcolor=blue]{hyperref}
\usepackage[pdftex,svgnames,usenames,dvipsnames]{xcolor}
\usepackage[utf8]{inputenc}
\usepackage[T1]{fontenc}
\usepackage{siunitx}
\usepackage{url}
\usepackage{newtxtext}
\usepackage[varg]{newtxmath}
\usepackage{bm}

\newcommand\figfolder{figs}

\begin{document}

\title{Shear thickening of suspensions of dimeric particles}

\author{Romain Mari}
\affiliation{Univ. Grenoble-Alpes, CNRS, LIPhy, 38000 Grenoble, France}

\date{\today}

\begin{abstract}
In this article, I study the shear thickening of suspensions of frictional dimers by the mean of numerical simulations.
I report the evolution of the main parameters of shear thickening, 
such as the jamming volume fractions in the unthickened and thickened branches of the flow curves, 
as a function of the aspect ratio of the dimers. 
The explored aspect ratios range from $1$ (spheres) to $2$ (dimers made of two kissing spheres).
I find a rheology qualitatively similar than the one for suspensions of spheres, except for the first normal stress difference $N_1$, 
which I systematically find negative for small asphericities.
I also investigate the orientational order of the particles under flow.
Overall, I find that dense suspensions of dimeric particles share many features with dry granular systems of elongated particles under shear, 
especially for the frictional state at large applied stresses. 
For the frictionless state at small stresses, I find that suspensions jam at lower volume fraction than dry systems, 
and that this difference increases with increasing aspect ratio. 
Moreover, in this state I find a thus far unobserved alignment of the dimers along the vorticity direction, 
as opposed to the commonly observed alignment with a direction close to the flow direction.
\end{abstract} 
\maketitle 

\section{Introduction}

Under shear, the viscosity of some dense suspensions of hard particles in the \SI{10}{\nano\meter} to \SI{100}{\micro\meter} range 
increases with the applied shear stress, a phenomenon known as shear thickening~\citep{Barnes_1989,Brown_2014,denn_rheology_2014}.
For volume fractions of solid material $\phi$ below a critical \(\phi_\mathrm{c}\), 
shear thickening is continuous (CST), that is, the viscosity $\eta$ 
is a continuous function of the shear rate \(\dot\gamma \), whereas for $\phi>\phi_\mathrm{c}$, 
shear thickening is discontinuous (DST), that is, the viscosity increases discontinuously at a given shear rate.

In the past few years, the frictional transition scenario was developed in order to explain shear thickening~\citep{morris_2018}.
In this scenario, shear thickening appears when two kinds of interparticles forces are present: 
a repulsive force (stemming from coated polymer brushes, electrical double layer, etc.)
and dry-like frictional forces, 
usually thought as a consequence of direct contact between particles 
by rupture of the lubrication film~\citep{fernandez_microscopic_2013,seto_discontinuous_2013,heussinger_shear_2013,lin_hydrodynamic_2015,royer_rheological_2016,clavaud_revealing_2017,comtet_pairwise_2017}.
At small applied stresses, because repulsive forces are able to maintain finite interparticle gaps, 
particles can move past each other under flow with a lubrication film separated them, 
making interactions virtually frictionless.
At large applied stresses on the other hand, repulsive forces are overcome and lubrication films are ruptured, 
giving way to frictional contacts between particles.
Because the viscosity divergence (i.e. the jamming transition) for systems made of frictionless particles occurs 
at a larger volume fraction than the viscosity divergence for systems made 
of frictional particles~\citep{silbert_geometry_2002,gallier_rheology_2014},
at a given $\phi$ the viscosity is larger in the large stress, frictional state than in the small stress, frictionless state.

\citet{Wyart_2014} introduced a scalar constitutive model for shear-thickening suspensions based on these ideas, 
relating the steady-state viscosity to the shear stress.
(Bashkirtseva et al.~\citep{bashkirtseva_rheophysics_2009} and Nakanishi et al.~\citep{Nakanishi_2012} also introduced 
similar types of constitutive model.)
In this model, the viscosity of the suspension diverges algebraically,
\(\eta  \sim  {\big(\phi_\mathrm{J}-\phi\big)}^{-\nu}\),
when approaching the jamming volume fraction \(\phi_\mathrm{J}\)~\citep{krieger_mechanism_1959,Zarraga_2000,Boyer_2011,Lerner_2012}. 
The value of the exponent \(\nu \) is debated~\citep{Lerner_2012,gallier_fictitious_2014,Mari_2014,ness_flow_2015}, 
the most typical value for spherical particles in the literature being \(\nu=2\)~\citep{ness_flow_2015,hermes_unsteady_2016,guy_constraint-based_2018}.
For long rods,~\citet{tapia_rheology_2017} measure  \(\nu=1\).
Shear thickening stems from the fact that $\phi_\mathrm{J}$, is at a higher value, $\phi_\mathrm{J}^0$, for frictionless particles than for frictional ones, at $\phi_\mathrm{J}^1 < \phi_\mathrm{J}^0$.
Because frictional contacts only exist at large stresses, 
due to the fact that under small stresses repulsive forces prevent them to form, 
the suspension has a stress dependent jamming point, $\phi_\mathrm{J}(\sigma)$, such that $\eta$ is an increasing function of $\sigma$.

This scenario is a priori independent of the particle shape.
While it has been tested mostly with suspensions of spherical particles, it is expected to be equally valid for 
suspensions of non-spherical particles, 
and many such suspensions are known to shear thicken in a qualitatively same way than spherical particles. 
Cornstarch suspension is the most famous example~\citep{Brown_2009,Fall_2012,oyarte_galvez_dramatic_2017}, 
but suspensions of synthetized particles were also studied~\citep{egres_rheology_2005,brown_shear_2011,royer_rheology_2015,james_controlling_2019,rathee_unraveling_2019}.
Most industrial suspensions known to shear thicken, 
such as cement paste~\citep{Lootens_2004,papo_effect_2004,feys_why_2009,toussaint_reducing_2009,roussel_steady_2010}, 
suspensions used for mechanical polishing such as fumed silica~\citep{crawford_shear_2012,amiri_stability_2012} or quartz powder suspensions~\citep{Freundlich_1938}, 
fresh paints and coatings~\citep{zupancic_rheological_1997,khandavalli_effect_2016}, 
or molten chocolate~\citep{blanco_conching_2019}, also contain particles of varied non-spherical shapes.
Numerical simulations of suspensions of frictional and repulsive aspherical particles also showed shear thickening~\citep{lorenz_combined_2018}.

The major difference with spherical particles comes from the different values for $\phi_\mathrm{J}^0$ and $\phi_\mathrm{J}^1$ 
(although other differences exist, e.g.~the exponent of the viscosity divergence close to jamming~\citep{tapia_rheology_2017}). 
Indeed, it is known that the jamming point for isotropic random packings generically depends on the shape of the particles~\citep{torquato_jammed_2010,jiao_maximally_2011,baule_fundamental_2014}.
For families of axisymmetric shapes (like axisymmetric ellipsoids, rods, spherocylinders, etc), which can be characterized by one scalar value, 
the aspect ratio $\alpha$ (the ratio between particle length and width), the behavior is non-monotonic as a function of $\alpha$.
Starting from spheres ($\alpha = 1$), the jamming volume fraction generally increases up to a maximum reached in the $\alpha = 1.2-2$ range, and then 
decreases with increasing $\alpha$ for larger aspect ratios~\citep{abreu_influence_2003,williams_random_2003,donev_improving_2004,donev_underconstrained_2007,jia_validation_2007,bargiel_geometrical_2008,wouterse_contact_2009,lu_computational_2010,kyrylyuk_isochoric_2011,zhao_dense_2012,saint-cyr_particle_2012,meng_shape_2012,ferreiro_cordova_random_2014,nan_dem_2015,meng_maximally_2016,vanderwerf_hypostatic_2018,marschall_compression_driven_2018,baule_mean_field_2013,kallus_random_2016}.

Morevover, in the case of elongated particles, the steady state under shear flow generally is not isotropic.
For dry granular systems of elongated particles, it is observed that the particles favor an orientation close to, 
but not quite the one of the flow direction~\citep{reddy_orientational_2009,campbell_elastic_2011,borzsonyi_orientational_2012,borzsonyi_shear_induced_2012,wegner_alignment_2012,guo_numerical_2012,guo_granular_2013,nagy_rheology_2017,trulsson_rheology_2018,nath_rheology_2018}.
A similar orientation is found in dense suspensions~\citep{egres_rheology_2005,rathee_unraveling_2019}, 
although the angle with the shear direction has not been systematically measured yet.
This anisotropy, introducing some degree of order, will also affect the volume fraction at which jamming occurs~\citep{wegner_alignment_2012,wegner_effects_2014,farhadi_dynamics_2014}.
Experiments in the literature were performed on suspensions of particles with rather large aspect ratios $\alpha > 2$, 
and found that, just like for isotropic systems, the jamming volume fraction is a decreasing function of $\alpha$~\citep{tapia_rheology_2017,bounoua_shear_2019,james_controlling_2019}.
For smaller aspect ratios, only numerical simulations for dry and inertial systems are available.
They show that for frictionless particles, the jamming volume fraction is typically larger than for isotropic systems. 
It increases at low $\alpha$, and then plateaus~\citep{trulsson_rheology_2018,nath_rheology_2018} or keeps increasing~\citep{nagy_rheology_2017} at large aspect ratios.
On the contrary, for large friction coefficients (typically $\mu_\mathrm{p} \gtrsim 1$), 
jamming occurs at volume fractions decreasing with $\alpha$ on the entire range of $\alpha$ studied~\citep{trulsson_rheology_2018,nath_rheology_2018}.
However, in all cases, for a given shape, the jamming volume fraction decreases with increasing friction coefficient, 
which is a sufficient condition to observe shear thickening for suspensions of repulsive particles.

In this article, I simulate dense suspensions of frictional and repulsive dimers, that is, 
particles made of two spheres connected by stiff springs approximating a rigid bond. 
I show that these systems shear thicken in the same way as systems of spherical particles, 
showing both continuous and discontinuous shear thickening when the volume fraction is varied.
I show results both for shear and normal stresses, revealing in particular that the 
first normal stress difference $N_1$ is always negative for particles with aspect ratios below $\alpha=2$.
I also study the orientational order of these suspensions, and uncover a previously unseen orientation 
in the frictionless state for small asphericities, with particles primarily aligning along the vorticity direction.

\section{Models and methods}

\subsection{General setup}
The simulation method for dimers is an adaptation of the one for spheres introduced in~\citet{seto_discontinuous_2013} 
and~\citet{Mari_2014}.
Dimers are built as pairs of spheres stuck together at a separation $s$ via a stiff spring and dashpot system.
As such, one can simulate a suspension of $N_\mathrm{D}$ dimers as a suspension of $N = 2N_\mathrm{D}$ spheres, 
only adding specific dimer interactions.
The aspect ratio $\alpha$ of a dimer is the ratio between its length and its width
\begin{equation}
  \alpha = \frac{a_i + a_j + s}{a_i + a_j} = 1+\frac{s}{a_i + a_j}.
\end{equation}
I here only consider dimers made of spheres $(i,\ j)$ with the same radii $a_i = a_j$, and such that they are connected, that is, $s\leq a_i+a_j$ 
and consequently $1\leq \alpha\leq 2$. 
Spheres have $\alpha = 1$, and most of the data I report here are for $\alpha=1.1,\ 1.2,\ 1.5$ and $2$.

I use a bidisperse mixture, with dimers made of spheres of radius either $a$ or $1.4a$.
The number ratio of two populations of dimers is such that each population occupies the same volume. 
The volume fraction, that is, the ratio between the volume occupied by the dimers and the total volume, is $\phi$.
The dimers are immersed in a Newtonian fluid with viscosity $\eta_\mathrm{f}$.
I consider the case of vanishing Stokes and Reynolds numbers (that is, inertialess fluid and particles).

The suspension rheology is studied under an imposed flow field $\bm{u}^{\infty}(\bm{r})$ characterized by 
a vorticity $\bm{\omega}^{\infty}$ and rate-of-strain tensor $\bm{e}^{\infty}$ as
\begin{equation}
  \bm{u}^{\infty}(\bm{r})
  =  \bm{\omega}^{\infty} \times \bm{r}
  + \bm{e}^{\infty} \cdot \bm{r}.
\end{equation}
This study focuses on simple shear flow, which corresponds to the
following nonzero elements: $\omega^{\infty}_3 = \dot{\gamma}/2$ and
$e^{\infty}_{12} = e^{\infty}_{21}= \dot{\gamma}/2$, 
where $\dot{\gamma}$ is the shear rate.
I use Lees-Edwards periodic boundary conditions to impose this deformation~\citep{Lees_1972}.

\subsection{Equations of motion}

The spherical particles are interacting through lubrication forces, contact forces, repulsive potential forces and dimer forces. 
At vanishing Stokes and Reynolds numbers, particles' equations of motion correspond to mechanical equilibrium 
on each particle in suspension, that is, a set of $6N$ coupled equations ($3N$ for force balance and $3N$ for force balance) in a three-dimensional system.
These equations can be compactly written as
\begin{equation}
 \bm{0} = 
 \begin{pmatrix}
    \bm{F}_\mathrm{H}\\
    \bm{T}_\mathrm{H}
  \end{pmatrix}
+
 \begin{pmatrix}
    \bm{F}_\mathrm{C}\\
    \bm{T}_\mathrm{C}
  \end{pmatrix}
+
 \begin{pmatrix}
    \bm{F}_\mathrm{D}\\
    \bm{T}_\mathrm{D}
  \end{pmatrix}
  +
 \begin{pmatrix}
    \bm{F}_\mathrm{R}\\
    \bm{0}
  \end{pmatrix}
,
  \label{eq:fb}
\end{equation}
where $\bm{F}_\mathrm{H} = \{\bm{f}_{\mathrm{H},1}, \dots, \bm{f}_{\mathrm{H},N}\}$ and 
$\bm{T}_\mathrm{H} = \{\bm{t}_{\mathrm{H},1}, \dots, \bm{t}_{\mathrm{H},N}\}$ are 
$3N$-dimensional hydrodynamic force (resp. hydrodynamic torque) 
vectors built from the $3$-dimensional force (resp. torque) vectors of every particle. 
Similarly, $\bm{F}_\mathrm{C}$ and $\bm{T}_\mathrm{C}$ contain 
the resultants of contact forces and torques on every particle, $\bm{F}_\mathrm{D}$ and $\bm{T}_\mathrm{D}$ contain 
the resultants of dimer forces and torques, $\bm{F}_\mathrm{R}$ contains the resultants of repulsive forces (which generate no torques).

\subsection{Forces}

All forces are pairwise (including hydrodynamic ones, as I only consider short-range lubrication), but dimer forces act only on pairs of spheres 
belonging to the same dimer, while repulsive, contact and hydrodynamic forces act only on pairs of spheres belonging to distinct dimers.

\subsubsection{Hydrodynamic forces}

Hydrodynamic forces and torques are coming from Stokes drag and lubrication forces between near particles, 
and are linearly related to the particles velocities as~\citep{Jeffrey_1984}
\begin{equation}
  \label{eq:f_hydro}
  \begin{pmatrix}
    \bm{F}_\mathrm{H}\\
    \bm{T}_\mathrm{H}
  \end{pmatrix} 
  = 
  -\bm{R}^{\mathrm{H}}_\mathrm{FU} \cdot
  \begin{pmatrix}
    \bm{U}-\bm{U}^{\infty}\\
    \bm{\Omega}- \bm{\Omega}^{\infty}
  \end{pmatrix}
  +
  \bm{R}^{\mathrm{H}}_\mathrm{FE}\vdots
  \bm{E}^{\infty}.
\end{equation}
The vectors $\bm{U} = \{\bm{u}_1, \dots, \bm{u}_N\}$ and $\bm{\Omega} = \{\bm{\omega}_1, \dots, \bm{\omega}_N\}$ are respectively the particle translational and angular velocities. 
Similarly, $\bm{U}^{\infty} = \{\bm{u}^{\infty}(\bm{r}_1), \dots, \bm{u}^{\infty}(\bm{r}_N)\}$, 
$\bm{\Omega}^{\infty} = \{\bm{\omega}^{\infty}(\bm{r}_1), \dots, \bm{\omega}^{\infty}(\bm{r}_N)\}$ and are the ``background'' imposed velocities 
(resp. angular velocities) evaluated at the particles' centers.
$\bm{E}^{\infty} = \{\bm{e}^{\infty}(\bm{r}_1), \dots, \bm{e}^{\infty}(\bm{r}_N)\}$ is a $N\times 3 \times 3$ tensor containing the ``background'' 
imposed rate-of-strain tensors for every particle.
Finally, $\bm{R}_\mathrm{FU}$ and $\bm{R}_\mathrm{FE}$ are resp. $6N\times 6N$ and $6N\times N\times 3\times 3$ resistance tensors, 
whose elements are given in detail in~\citep{Mari_2014}. 
They include the leading order of lubrication terms as defined in~\citet{Jeffrey_1984} 
that are diverging when the normalized separation gap 
$h^{(i,j)} = 2(|\bm{r}_j - \bm{r}_i| - a_i - a_j)/(a_i + a_j)$ vanishes; 
they correspond to the ``squeeze'', ``shear'' and ``pump'' modes of~\citet{Ball_1997}.
While these terms diverge as either $1/h^{(i,j)}$ or $\log h^{(i,j)}$ for ideally smooth spheres,
they are regularized by introducing a roughness length $\delta = 10^{-2}$ such that they scale 
respectively as $1/(h^{(i,j)}+\delta)$ or $\log(h^{(i,j)}+\delta)$.
Note that in $\bm{R}^{\mathrm{H}}_\mathrm{FU} \cdot (\bm{U}-\bm{U}^{\infty}, \bm{\Omega}- \bm{\Omega}^{\infty})$ 
the contraction is over the second index of $\bm{R}_\mathrm{FU}$, 
while in $\bm{R}^{\mathrm{H}}_\mathrm{FE}\vdots\bm{E}^{\infty}$ it is on the last three indices of $\bm{R}^{\mathrm{H}}_\mathrm{FE}$.
I consider that two particles $i$ and $j$ exchange hydrodynamic forces only if their normalized separation gap 
is such that $0<h^{(i,j)}<0.2$.

\subsubsection{Contact forces}

The contact force on a particle \(i\) with radius \(a_i\) in contact with particle \(j\)
can be decomposed in normal and tangential components
\begin{equation}
  \bm{f}_{\mathrm{C}}^{(i,j)} = \bm{f}_{\mathrm{C,nor}}^{(i,j)} + \bm{f}_{\mathrm{C, tan}}^{(i,j)}.
\end{equation}
Contacts fulfill Coulomb's friction laws 
\(\bigl|\bm{f}_{\mathrm{C,tan}}^{(i,j)} \bigr| \leq \mu_\mathrm{p} |\bm{f}_{\mathrm{C,nor}}^{(i,j)}|\) with 
sliding friction coefficient \( \mu_\mathrm{p} \).
The force components \(\bm{f}_{\mathrm{C,nor}}^{(i,j)}\) and \(\bm{f}_{\mathrm{C, tan}}^{(i,j)}\) 
are modeled in a Cundall-Strack fashion~\citep{Cundall_1979, Luding_2008}, with normal and tangential couples of spring and dashpot
\begin{equation}
  \label{eq:contact_model}
  \begin{split}
    \bm{f}_{\mathrm{C,nor}}^{(i,j)} & = k_{\mathrm{n}} \bm{\xi}_{\mathrm{C,n}}^{(i,j)} 
     + \gamma_{\mathrm{n}}  \bm{u}_{\mathrm{n}}^{(i,j)},  \\
    \bm{f}_{\mathrm{C,tan}}^{(i,j)} & = k_{\mathrm{t}} \bm{\xi}_{\mathrm{C,t}}^{(i,j)}
    +\gamma_{\mathrm{t}}  \bm{u}_{\mathrm{t}}^{(i,j)},
  \end{split}
\end{equation}
where $\bm{n}_{ij} = (\bm{r}_j - \bm{r}_i)/|\bm{r}_j - \bm{r}_i|$ is the unit center-to-center vector, 
$\bm{\xi}_{\mathrm{C,n}}^{(i,j)} = h^{(i,j)} \bm{n}_{ij}$ is the normal spring stretch, 
$\bm{\xi}_{\mathrm{C,t}}^{(i,j)}$ is the tangential spring stretch,
$\bm{u}_{\mathrm{n}}^{(i,j)} = (\bm{I} - \bm{n}_{ij}\bm{n}_{ij})\cdot (\bm{u}_j - \bm{u}_i)$ is the normal velocity difference (with $\bm{I}$ the identity matrix),
and $\bm{u}_{\mathrm{t}}^{(i,j)} = \bm{u}_j - \bm{u}_i - \bm{u}_{\mathrm{n}}^{(i,j)} - (a_i \bm{\omega}_i + a_j \bm{\omega}_j)\times \bm{n}_{ij}$ is the tangential relative surface velocity.
Finally, the contact torque on particle $i$ from the contact with particle $j$ is simply obtained as 
$\bm{t}_{\mathrm{C}}^{(i,j)} = a_i \bm{n}_{ij} \times \bm{f}_{\mathrm{C}}^{(i,j)}$. Note that a contact is active only if $h^{(i,j)} < 0$.
Contact forces and torques are thus decomposed in velocity-independent (from springs) and velocity-proportional (from dashpots) parts
\begin{equation}
  \label{eq:f_contact}
  \begin{pmatrix}
    \bm{F}_\mathrm{C}\\
    \bm{T}_\mathrm{C}
  \end{pmatrix}
  =
  \begin{pmatrix}
    \bm{F}_\mathrm{C,Spring}\\
    \bm{T}_\mathrm{C,Spring}
  \end{pmatrix}
  -
  \bm{R}^{\mathrm{C}}_\mathrm{FU} \cdot
  \begin{pmatrix}
    \bm{U}\\
    \bm{\Omega}
  \end{pmatrix}
\end{equation}
with $\bm{R}^{\mathrm{C}}_\mathrm{FU}$ the dashpot resistance matrix.

\subsubsection{Dimer forces}

For a rigid object, all points $\bm{r}$ inside the object have the same angular velocity $\bm{\omega}(\bm{r})=\bar{\bm{\omega}}$, 
and translational velocities such that, if $\bm{r}'$ is any other point in the object, 
$\bm{u}(\bm{r}) = \bm{u}(\bm{r}') + \bar{\bm{\omega}}\times (\bm{r} - \bm{r}')$.
In this work, I want to simulate a rigid dimer as an object made of two independent rigid spheres. 
The force and torque coupling the two spheres (say, $i$ and $j$) of a dimer are also modeled in Cundall-Strack manner, 
with two couples of spring and dashpot (a similar technique, only without dashpot, was already adopted by~\citet{yamamoto_method_1993,yamamoto_viscosity_1994}). 
Just like in the contact case, the springs provide restoring forces and torques when a deformation of the dimer occurs, 
and the dashpots ensure that relaxations towards the undeformed state are slow enough to not require unreasonably small time steps 
to resolve accurately, but also fast enough compared to the typical physical processes to be captured, in the present case occuring on a timescale $\propto \dot\gamma^{-1}$.
The first spring-dashpot system keeps particles angular velocities as close as possible, that is, minimizes $\delta \bm{\omega}^{(i,j)} = \bm{\omega_j} - \bm{\omega_j}$ 
and gives rise to a torque
\begin{equation}
  \label{eq:f_dimer_angular}
\bm{t}_{\mathrm{D,a}}^{(i,j)} = k_{\mathrm{D,a}} \bm{\xi}_{\mathrm{D,a}} + \gamma_{\mathrm{D,a}}\delta \bm{\omega}^{(i,j)},
\end{equation}
with $k_{\mathrm{D,a}}$ the spring stiffness, $\gamma_{\mathrm{D,a}}$ the dashpot resistance, 
and $\bm{\xi}_{\mathrm{D,a}}$ the spring stretch computed as the cumulated rotational displacement, 
that is, $\mathrm{d}\bm{\xi}_{\mathrm{D,a}}/\mathrm{d}t = \delta \bm{\omega}^{(i,j)}$. 
The second spring-dashpot system ensures that the translational velocities of the two spheres are as close as possible 
from the one of a rigid dimer with angular velocity $(\bm{\omega_j} + \bm{\omega_j})/2$, and generates a force and a torque
\begin{equation}
  \label{eq:f_dimer_translational}
  \begin{split}
    \bm{f}_{\mathrm{D,t}}^{(i,j)} & = k_{\mathrm{D,t}} \bm{\xi}_{\mathrm{D,t}} + \gamma_{\mathrm{D,t}} \bm{u}^{(i,j)}_{\mathrm{D,t}}, \\
    \bm{t}_{\mathrm{D,t}}^{(i,j)} & = \frac{(\bm{r}_j - \bm{r}_i)}{2}\times \bm{f}_{\mathrm{D,t}}^{(i,j)},
  \end{split}
\end{equation}
with $\bm{u}^{(i,j)}_{\mathrm{D,t}} = \bm{u}_j - \bm{u}_i - (\bm{\omega_j} + \bm{\omega_j}) \times (\bm{r}_j - \bm{r}_i)/2$.
It then appears that the dimer forces and torques can also be expressed as the sum of a velocity-independent spring term, and a velocity-proportional dashpot term
\begin{equation}
  \label{eq:f_dimer}
  \begin{pmatrix}
    \bm{F}_\mathrm{D}\\
    \bm{T}_\mathrm{D}
  \end{pmatrix}
  =
  \begin{pmatrix}
    \bm{F}_\mathrm{D,Spring}\\
    \bm{T}_\mathrm{D,Spring}
  \end{pmatrix}
  -
  \bm{R}^{\mathrm{D}}_\mathrm{FU} \cdot
  \begin{pmatrix}
    \bm{U}\\
    \bm{\Omega}
  \end{pmatrix}
\end{equation}

\subsubsection{Repulsive forces}

The repulsive forces are exponentially decaying with the gap on a typical lengthscale $\lambda$
\begin{equation}
  \label{eq:f_repulsion}
    \bm{f}_{\mathrm{R}}^{(i,j)}  = - A_{\mathrm{R}} \exp\left(-\frac{h^{(i,j)}}{\lambda}\right) \bm{n}^{(i,j)}.
\end{equation}
In this work I use $\lambda = 0.02$, and in order to keep the computational time low I neglect 
the repulsive force if $h^{(i,j)}>7\lambda$. Importantly, this force defines the stress unit scale
\begin{equation}
    \sigma_{\mathrm{r}}  = \frac{A_{\mathrm{R}}}{a^2},
\end{equation}
which is the typical stress scale around which shear thickening occurs.

\subsection{Stresses}

The total stress of the dimer suspension is the sum of the solvent stress and the particle stress, itself a sum of individual dimer contributions
\begin{equation}
\Sigma = 2 \eta_\mathrm{f} \bm{e}^{\infty} + V^{-1}\sum_\alpha \bm{\sigma}_\alpha
\end{equation}
with $V$ the total volume of the system.

The stresslet $\bm{\sigma}_\alpha$ is the symmetrized first moments of the force field $\bm{f}_\alpha$ acting on the surface $\partial \alpha$ of dimer $\alpha$
\begin{equation}
 \bm{\sigma}_\alpha = - \frac{1}{2} \int_{\partial \alpha} \mathrm{d}\bm{r} \left[(\bm{r} - \bm{r}_\alpha) \bm{f}_\alpha(\bm{r}) + \bm{f}_\alpha(\bm{r})(\bm{r} - \bm{r}_\alpha)\right].
 \end{equation}
Note that because the dimers are force free, the stresslet does not depend on the ``reference point'' $\bm{r}_\alpha$. 
For the same reason, the stresslet can be further decomposed in contributions from the two spheres $i$ and $j$ making the dimer
\begin{equation}
 \bm{\sigma}_\alpha = \bm{\sigma}_i + \bm{\sigma}_j = - \frac{1}{2} \int_{\partial i} \mathrm{d}\bm{r} \left[(\bm{r} - \bm{r}_i) \bm{f}_i(\bm{r}) + \bm{f}_i(\bm{r})(\bm{r} - \bm{r}_i)\right]
                                        - \frac{1}{2} \int_{\partial j} \mathrm{d}\bm{r} \left[(\bm{r} - \bm{r}_j) \bm{f}_j(\bm{r}) + \bm{f}_j(\bm{r})(\bm{r} - \bm{r}_j)\right],
 \end{equation}
where the force densities $\bm{f}_i$ and $\bm{f}_j$ now include the dimer forces, and $\partial i$ and $\partial j$ are the sphere surfaces truncated at the dimer mid-plane.
Again, I conventionally take $\bm{r}_i$ as the center of sphere $i$ (and similarly for $\bm{r}_j$), but the stresslets do not depend on this choice.

The individual stresslets have contributions from hydrodynamic, contact, repulsion and dimer forces. 
Hydrodynamic stresslets can be written with the compact notation $\bm{S}_{\mathrm{H}} = \{\bm{\sigma}_{\mathrm{H},1}, \dots, \bm{\sigma}_{\mathrm{H},N}\}$ 
used when introducing forces, as~\citep{Jeffrey_1992}
\begin{equation}
\label{eq:hydro_stress}
  \bm{S}_{\mathrm{H}} =
  -
  \bm{R}_{\mathrm{SU}} \cdot 
  \begin{pmatrix}
    \bm{U} - \bm{U}^{\infty} \\
    \bm{\Omega} - \bm{\Omega}^{\infty}
    \end{pmatrix}
    +
    \bm{R}_{\mathrm{SE}}\vdots\bm{E}^{\infty},
\end{equation}
where $\bm{R}_{\mathrm{SU}}$ and $\bm{R}_{\mathrm{SE}}$ are resistance tensors from the lubrication forces considered~\citep{Mari_2014}.

For all other forces, the stresslets are sums of contributions from individual interactions, and are written as 
($\bullet$ standing for $\mathrm{C}$, $\mathrm{R}$ or $\mathrm{D}$)
\begin{equation}
  \bm{\sigma}_{\bullet, i} = - \frac{1}{2} \sum_j \frac{a_i}{a_i+a_j} \left[ (\bm{r}_j - \bm{r}_i) \bm{f}_{\bullet}^{(i,j)} + \bm{f}_{\bullet}^{(i,j)} (\bm{r}_j - \bm{r}_i)\right].
\end{equation}
which defines a linear operator $\bm{X}$ such that
\begin{equation}
\label{eq:xf_stress}
  \bm{S}_{\bullet} = \bm{X} \cdot \mathcal{F}_{\bullet}
\end{equation}
with $\bm{S}_{\bullet} = \{\bm{\sigma}_{\bullet,1}, \dots, \bm{\sigma}_{\bullet,N}\}$ and $\mathcal{\bm{F}}_{\bullet} = \{\bm{f}_{\bullet}^{(i, j)}\}_{\{(i,j)\}}$.

\subsection{Controlled stress algorithm}
One can solve these equations of motion under constant shear rate 
or under constant shear stress. 
Constant shear stress simulations are better suited to the study of abrupt shear thickening than constant shear rate simulations, as 
the former show much smaller viscosity fluctuations than the latter in the shear thickening regime~\citep{mari_nonmonotonic_2015}. 
This is fundamentally rooted in the stress controlled nature of shear thickening~\citep{Brown_2014}.
Moreover, experimental rheometry on shear thickening suspensions is more commonly performed in controlled-stress conditions.
Therefore, in this work, I chose constant stress conditions, using the algorithm of~\citet{mari_nonmonotonic_2015}. 
Using Eqs.~\ref{eq:f_hydro}, \ref{eq:f_contact} and \ref{eq:f_dimer}, one can rewrite the equations of motion Eq.~\ref{eq:fb} as
\begin{equation}
 \bm{0} = 
 - \bm{R}_\mathrm{FU}
 \cdot
  \begin{pmatrix}
    \bm{U}-\bm{U}^{\infty}\\
    \bm{\Omega}- \bm{\Omega}^{\infty}
  \end{pmatrix}
  +
  \bm{R}^{\mathrm{H}}_\mathrm{FE}\vdots
  \bm{E}^{\infty}
+
 \begin{pmatrix}
    \bm{F}_\mathrm{C,Spring}\\
    \bm{T}_\mathrm{C,Spring}
  \end{pmatrix}
+
 \begin{pmatrix}
    \bm{F}_\mathrm{D,Spring}\\
    \bm{T}_\mathrm{D,Spring}
  \end{pmatrix}
  +
 \begin{pmatrix}
    \bm{F}_\mathrm{R}\\
    \bm{0}
  \end{pmatrix}
  +
  \left[ \bm{R}^{\mathrm{C}}_\mathrm{FU} + \bm{R}^{\mathrm{D}}_\mathrm{FU} \right]
 \cdot
  \begin{pmatrix}
    \bm{U}^{\infty}\\
    \bm{\Omega}^{\infty}
  \end{pmatrix}
.
  \label{eq:eq_motion}
\end{equation}
with $\bm{R}_\mathrm{FU} = \bm{R}^{\mathrm{H}}_\mathrm{FU} + \bm{R}^{\mathrm{C}}_\mathrm{FU} + \bm{R}^{\mathrm{D}}_\mathrm{FU}$.

Similarly, using Eqs.~\ref{eq:hydro_stress} and \ref{eq:xf_stress}, 
the total stress $\bm{S} = \bm{S}_{\mathrm{H}} + \bm{S}_{\mathrm{C}} + \bm{S}_{\mathrm{D}} + \bm{S}_{\mathrm{R}}$ is 
\begin{equation}
\bm{S} = 
-
  \bm{R}_{\mathrm{SU}} \cdot 
  \begin{pmatrix}
    \bm{U} - \bm{U}^{\infty} \\
    \bm{\Omega} - \bm{\Omega}^{\infty}
    \end{pmatrix}
    +
    \bm{R}_{\mathrm{SE}}\vdots\bm{E}^{\infty}
    +
    \bm{X}.\mathcal{F}_{\mathrm{C}}
    +
    \bm{X}.\mathcal{F}_{\mathrm{D}}
    +
    \bm{X}.\mathcal{F}_{\mathrm{R}}
\end{equation}
Maintaining the shear stress at a constant value $\tau$ means that at any time $t$ in the simulation one has to determine 
the shear rate $\dot\gamma(t)$ such that
\begin{equation}
  \Sigma_{12} = \sum_i (\bm{\sigma}_i)_{12} = \tau
\end{equation}
In the present case, this is easily achievable because $\bm{\Sigma}$ is an affine function of $\dot\gamma$.
Indeed, the solution to Eq.~\ref{eq:eq_motion} is itself affine
\begin{equation}
\begin{split}
\begin{pmatrix}
    \bm{U}\\
    \bm{\Omega}
  \end{pmatrix}
  & =
  \begin{pmatrix}
    \bm{U}^{\infty}\\
    \bm{\Omega}^{\infty}
  \end{pmatrix}
  + 
\bm{R}_\mathrm{FU}^{-1} \cdot 
\bigg[ 
\bm{R}^{\mathrm{H}}_\mathrm{FE}\vdots
  \bm{E}^{\infty}
+
 \begin{pmatrix}
    \bm{F}_\mathrm{C,Spring}\\
    \bm{T}_\mathrm{C,Spring}
  \end{pmatrix}
+
 \begin{pmatrix}
    \bm{F}_\mathrm{D,Spring}\\
    \bm{T}_\mathrm{D,Spring}
  \end{pmatrix}
  +
 \begin{pmatrix}
    \bm{F}_\mathrm{R}\\
    \bm{0}
  \end{pmatrix}
  +
  \left[ \bm{R}^{\mathrm{C}}_\mathrm{FU} + \bm{R}^{\mathrm{D}}_\mathrm{FU} \right]
 \cdot
  \begin{pmatrix}
    \bm{U}^{\infty}\\
    \bm{\Omega}^{\infty}
  \end{pmatrix}
  \bigg] \\
& \equiv 
\dot\gamma
\begin{pmatrix}
    \hat{\bm{U}}_\mathrm{prop}\\
    \hat{\bm{\Omega}}_\mathrm{prop}
  \end{pmatrix}
  +
  \begin{pmatrix}
    \bm{U}_\mathrm{indep}\\
    \bm{\Omega}_\mathrm{indep}
  \end{pmatrix}
\end{split}
\label{eq:vel_split}
\end{equation}
with the two shear-rate independent factors
\begin{align}
\begin{pmatrix}
    \hat{\bm{U}}_\mathrm{prop}\\
    \hat{\bm{\Omega}}_\mathrm{prop}
  \end{pmatrix}
 & =
\begin{pmatrix}
    \hat{\bm{U}}^{\infty}\\
    \hat{\bm{\Omega}}^{\infty}
  \end{pmatrix}
  + 
\bm{R}_\mathrm{FU}^{-1} \cdot 
\bigg[ 
\bm{R}^{\mathrm{H}}_\mathrm{FE}\vdots
  \hat{\bm{E}}^{\infty}
  +
  \left[ \bm{R}^{\mathrm{C}}_\mathrm{FU} + \bm{R}^{\mathrm{D}}_\mathrm{FU} \right]
 \cdot
  \begin{pmatrix}
    \hat{\bm{U}}^{\infty}\\
    \hat{\bm{\Omega}}^{\infty}
  \end{pmatrix}
  \bigg], \\
\begin{pmatrix}
    \bm{U}_\mathrm{indep}\\
    \bm{\Omega}_\mathrm{indep}
  \end{pmatrix}
  &=
  \bm{R}_\mathrm{FU}^{-1} \cdot 
\bigg[ 
 \begin{pmatrix}
    \bm{F}_\mathrm{C,Spring}\\
    \bm{T}_\mathrm{C,Spring}
  \end{pmatrix}
+
 \begin{pmatrix}
    \bm{F}_\mathrm{D,Spring}\\
    \bm{T}_\mathrm{D,Spring}
  \end{pmatrix}
  +
 \begin{pmatrix}
    \bm{F}_\mathrm{R}\\
    \bm{0}
  \end{pmatrix}
  \bigg],
\end{align}
where $\hat{\bm{U}}^{\infty} = \bm{U}^{\infty}/\dot\gamma$, 
$\hat{\bm{\Omega}}^{\infty} = \bm{\Omega}^{\infty}/\dot\gamma$ and $\hat{\bm{E}}^{\infty} = \bm{E}^{\infty}/\dot\gamma$.
As a consequence, the stress is affine
\begin{equation}
\begin{split}
\bm{S} &= 
\dot\gamma \bigg\{
\bm{R}_{\mathrm{SU}} \cdot 
  \begin{pmatrix}
    \hat{\bm{U}}_\mathrm{prop} - \hat{\bm{U}}^{\infty} \\
    \hat{\bm{\Omega}}_\mathrm{prop} - \hat{\bm{\Omega}}^{\infty}
  \end{pmatrix}
    +
    \bm{R}_{\mathrm{SE}}\vdots\hat{\bm{E}}^{\infty}
    \bigg\}
    +
    \bm{R}_{\mathrm{SU}} \cdot 
  \begin{pmatrix}
    \bm{U}_\mathrm{indep} \\
    \bm{\Omega}_\mathrm{indep}
  \end{pmatrix}
+
    \bm{X}.\mathcal{F}_{\mathrm{C}}
    +
    \bm{X}.\mathcal{F}_{\mathrm{D}}
    +
    \bm{X}.\mathcal{F}_{\mathrm{R}} \\
    & \equiv
    \dot\gamma \hat{\bm{S}}_\mathrm{prop} + \bm{S}_\mathrm{indep}.
    \end{split}
    \label{eq:stress_split}
\end{equation}
Note that even if at each time step, the stress is an affine function of the shear rate, 
this does not imply that the steady-state rheology is affine.
Indeed, steady states at different values of stress $\tau$ 
(or, equivalently, different values of $\dot\gamma$) have markedly different microstructures, 
with the large $\tau$ states having many contacts, whereas there are none at low stresses. 
This comes from the different balance between the terms involved in the equation of motion~\ref{eq:eq_motion}, 
and ultimately dramatically affects the weights of $\hat{\bm{S}}_\mathrm{prop}$ and $\bm{S}_\mathrm{indep}$ in Eq.~\ref{eq:stress_split}, 
which are far from independent of $\dot\gamma$ when averaged over steady states.

Eventually, at each time step one first computes the shear rate as
\begin{equation}
\dot\gamma = \frac{\tau - \sum_i \left(\bm{\sigma}_\mathrm{indep}\right)_{12}}{\sum_i \left(\hat{\bm{\sigma}}_\mathrm{prop}\right)_{12}}
\end{equation}
and then compute the velocities with Eq.~\ref{eq:vel_split}.
From the velocities, one obtains the positions $\bm{r}(t+dt)$ of the particles in time $t+dt$
from the positions $\bm{r}(t)$ at time $t$ through time integration, with a predictor-corrector algorithm.
I used an adaptative time step $dt$ ensuring that, with a dimensionless coefficient $\epsilon = 5\times 10^{-4}$, 
for any pairs of particles $i, j$ in interaction, $\max(\bm{u}_{\mathrm{n}}^{(i,j)}, \bm{u}_{\mathrm{t}}^{(i,j)}, \bm{u}_{\mathrm{D,t}}^{(i,j)}) \leq \epsilon(a_i+a_j)/(2dt)$ 
and $\delta \bm{\omega}^{(i,j)} \leq \epsilon/dt$.

\subsection{Choice of parameters}

The numerical model contains a few free parameters. 
In particular, the several springs and dashpots involved in the dimer and contact model need to be picked carefully.

The tradeoff for these parameters is always the same. 
Spring constants should be large enough to be in the hard and rigid particle limit. 
This is expected when the deformation from the applied stress are on a scale much smaller than the particle size, 
that is, when $\tau$ satisfies $\tau \ll k_{\mathrm{n}}/a,\ k_{\mathrm{t}}/a,\ k_{\mathrm{D,a}}/a^2,\ k_{\mathrm{D,t}}/a$.
However, one has to resolve particle trajectories with a resolution such that at each time step the particles displacements 
are only a fraction of the spring stretches, otherwise the spring forces are not resolved in a smooth manner, which generate numerical instabilities. 
Because larger stiffnesses allow smaller spring stretches, there is a limit to the increase of stiffnesses beyond which the required 
time stepping make simulation times prohibitive.
This can be mitigated by the use of larger dashpot resistances, which slow spring relaxations, but as explained earlier the spring relaxation times 
have to stay much smaller than the physical timescales of the phenomena one wants to simulate.
As a consequence of this tradeoff, in this work I chose $k_{\mathrm{n}} = 5\times 10^3\tau a$, $k_{\mathrm{t}} = 2.5\times 10^3\tau a$. 
With this choice, the minimum interparticle gap between any pair of particles during the simulation is maintained at $\min_{i,j} h^{(i,j)} \approx -0.02$.

With the same motivation, I used $k_{\mathrm{D,a}} = 10^4\tau a^2$ and $k_{\mathrm{D,t}} = 10^4\tau a$. 
To check the actual deformation of the dimers, one can for instance measure the actual aspect ratios of the dimers during the simulation, 
in particular at the largest stress values.
The distribution $\rho(\alpha)$ obtained for $\phi = 0.5$ and $\alpha=0.5$ at the largest simulated stress $\tau/\sigma_\mathrm{r} = 100$, 
shown in green line in the inset of Fig.~\ref{fig:rheology_rate}, 
confirms that all dimers keep an aspect ratio within $\SI{3}{\percent}$ (and for more than $\SI{80}{\percent}$ of them within $\SI{1}{\percent}$) of the nominal value.
While this distribution can be further narrowed by doubling the stiffnesses to $k_{\mathrm{D,a}} = 2\times 10^4\tau a^2$ and $k_{\mathrm{D,t}} = 2\times 10^4\tau a$ (black line), 
it is not necessary, as the system already is in the rigid dimer limit in which the rheology does not depend on the stiffness values anymore.
I indeed show in Fig.~\ref{fig:rheology_rate} that the viscosity measured with the regular dimer spring stiffnesses is undistinguishable 
from the the one measured with twice larger stiffnesses. 

The data in this article are obtained with $N_\mathrm{D} = 500$ dimers. 
Data shown are obtained with a friction coefficient $\mu_\mathrm{p} = 0.5$, 
unless stated otherwise. 
I varied systematically aspect ratio, applied stress and volume fraction. 
For each combination of these three parameters, data averages are taken over a single simulation of at least 15 strain units (and up to 50 strain units), 
after discarding the first 5 strain units to avoid averaging over the start-up transients.
Initial configurations are generated as follows. 
I start by placing $N_\mathrm{D}$ spheres in the simulation box with positions picked at random in a uniform distribution. 
Dimers are created by duplicating each sphere with a twin sphere placed at a shifted position separated from the initial sphere position 
by a distance corresponding to the nominal aspect ratio of the dimer along a direction picked at random with a uniform distribution on the unit sphere, 
taking properly into account periodic boundary conditions.
At the volume fractions considered in this article, these random configurations are typically giving highly overlaped particles. 
To obtain actual intial configurations, I therefore let these random configurations relax to overlap-free configurations (which is always possible below the jamming transition) 
by running the algorithm described in this section, only for frictionless dimers and in controlled-rate conditions with a vanishing shear rate.
 
\section{Results}

\subsection{Shear viscosity}
\begin{figure}[t]
  \centering
  \includegraphics[width=0.9\textwidth]{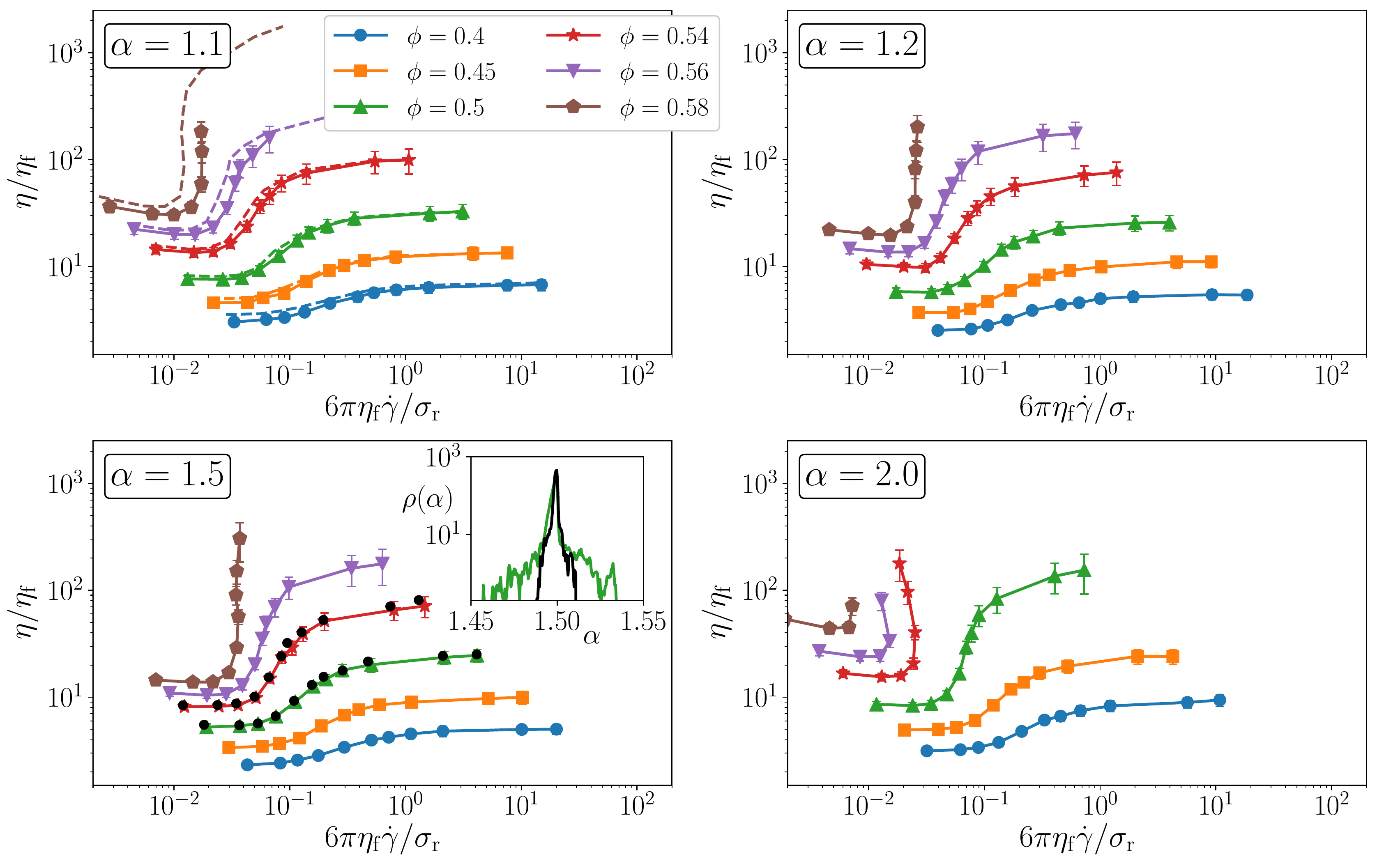}
  \caption{Relative viscosity $\eta/\eta_\mathrm{f}$ as a function of the shear rate for spheres (aspect ratio $\alpha=1$, top left, dashed lines), and for dimers with aspect ratios  $\alpha=1.1$ (top left, symbols), $\alpha=1.2$ (top right), $\alpha=1.5$ (bottom left, colored symbols) and $\alpha=1.2$ (bottom right). Volume fractions $\phi=0.4,\ 0.45,\ 0.5,\ 0.54,\ 0.56$ and $0.58$ are reported.
  The error bars represent the standard deviation observed in the time series of the viscosity in steady state.
  In the bottom left, I show in black circles 
  the viscosities obtained for $\phi=0.5$ and $\phi=0.54$ with twice larger value for the stiffnesses of the dimer spring $k_{\mathrm{D,a}}$ and $k_{\mathrm{D,t}}$.
  For the $\phi=0.5$, $\alpha=1.5$ data, I also show in the inset the distribution of measured $\alpha$ during the simulation, for both the ``regular'' stiffness (green) and the double stiffness (black).
  }\label{fig:rheology_rate}
\end{figure}

I first show the flow curves $\eta(\dot\gamma)$ obtained at several volume fractions, 
for aspect ratios $\alpha = 1.1, 1.2, 1.5$ and $2$ 
in Fig.~\ref{fig:rheology_rate}. Spheres ($\alpha = 1$) are also shown, for comparison.
These flow curves are typical of a shear thickening suspension, and follow the same qualitative behaviour than for a suspension of spheres.
Shear thickening occurs between a contactless, frictionless state at low stresses ($\tau\ll\sigma_\mathrm{r}$), 
and a frictional state where contacts proliferate at high stresses ($\tau\gg \sigma_\mathrm{r}$).
At low volume fractions, shear thickening is continuous.
Above a critical volume fraction $\phi_\mathrm{c}(\alpha)$, shear thickening is discontinuous, 
a situation characterized by a S-shaped flow curve 
in Fig.~\ref{fig:rheology_rate}.

The overall values of the viscosity depend on the aspect ratio, but in a non-monotonic way: 
while the viscosity at a given volume fraction and applied stress decreases when $\alpha$ increases from $1.$ to $1.5$, 
it increases between $\alpha=1.5$ and $\alpha=2$. 
The aspect ratio also influences $\phi_\mathrm{c}(\alpha)$, which is roughly located around $0.58$ for $\alpha=1.1$, $\alpha=1.2$, 
and $\alpha=1.5$, but is at a much lower value between $\phi=0.5$ and $\phi=0.54$ for $\alpha=2$.

\begin{figure}[t]
  \centering
  \includegraphics[width=0.9\textwidth]{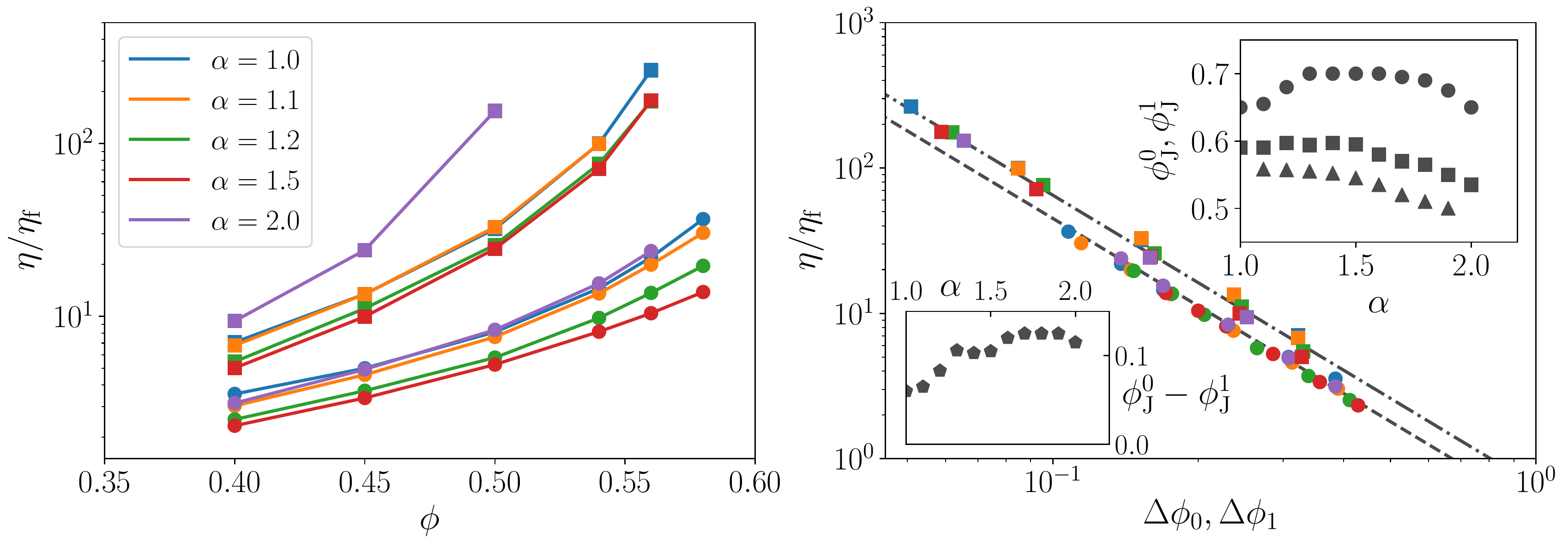}
  \caption{\textbf{Left:} Minima $\eta_\mathrm{min}$ (circles) and maxima $\eta_\mathrm{max}$ (squares) 
  of the flow curves in Fig.~\ref{fig:rheology_rate} as a function of the volume fraction $\phi$, for aspect ratios $\alpha=1, 1.1, 1.2, 1.5$ and $2$. 
  \textbf{Right:} Same data, plotted in a logarithmic scale as a function of $\Delta\phi_0 = \phi_\mathrm{J}^0 - \phi$ 
  in the case of $\eta_\mathrm{min}$, and $\Delta\phi_1 = \phi_\mathrm{J}^1 - \phi$ in the case of $\eta_\mathrm{max}$. 
  I show the best fits to $\eta_\mathrm{min} \sim \Delta \phi_0^{-2}$ and  $\eta_\mathrm{max} \sim \Delta \phi_1^{-2}$, 
  which determine values of $\phi_\mathrm{J}^0$ and $\phi_\mathrm{J}^1$ for each $\alpha$. 
  I show the obtained values for $\phi_\mathrm{J}^0$ (circles), $\phi_\mathrm{J}^1$ for $\mu_{\mathrm{p}} = 0.5$ (squares), 
  and also $\phi_\mathrm{J}^1$ for $\mu_{\mathrm{p}} = 1$ (triangles) for more values of $\alpha$, in the top right inset. In the bottom left inset the difference $\phi_\mathrm{J}^0- \phi_\mathrm{J}^1$ as a function of $\alpha$, for the data with $\mu_{\mathrm{p}} = 0.5$.}\label{fig:jamming}
\end{figure}

This can be rationalized by the $\alpha$ dependence of the frictionless and frictional jamming volume fractions, respectively
$\phi_\mathrm{J}^0$ and $\phi_\mathrm{J}^1$.
I evaluate $\phi_\mathrm{J}^0$ by tracking the viscosity minimum $\eta_\mathrm{min}$ at low shear stresses as a function of $\phi$ and 
fitting its divergence to $\eta_\mathrm{min} \sim (\phi_\mathrm{J}^0 - \phi)^{-\nu} \equiv \Delta \phi_0^{-\nu}$, 
as shown in Fig.~\ref{fig:jamming}. 
For simplicity I here make the common choice of an exponent $\nu = 2$, 
and as a consequence focus rather on the trend for $\phi_\mathrm{J}^0$ 
(which should be less affected by possible variations of the actual exponent) more than its absolute value.
(However, note that the data cannot be reasonably fit by exponents very different from $\nu=2$, and in particular data are clearly incompatible 
with $\nu=1$ even for the largest aspect ratio studied here, $\alpha=2$, which contrasts with the situation experimentally observed for $\alpha\gg 1$~\citep{tapia_rheology_2017}.)
I find that $\phi_\mathrm{J}^0$ first increases as a function of $\alpha$, from $\alpha=1$ to $\alpha \approx 1.2$, followed by a broad maximum from  $\alpha \approx 1.2$ to $\alpha \approx 1.5$, and then decreases for larger $\alpha$.
In contrast to earlier numerical simulations 
of dry granular systems~\citep{nagy_rheology_2017,nath_rheology_2018,trulsson_rheology_2018}, 
I do not find a plateau for $\alpha \gtrsim 1.5$.
In the dry case, this plateau was observed only for frictionless particles and was attributed to the rather strong ordering under shear. 
It is likely that for the present suspensions, lubrication plays a similar role to friction in creating torques that destabilize particles alignment.
Indeed, I will show later on that the degree of order observed in my simulations is smaller 
than what was observed in dry granular systems, whether in two~\citep{trulsson_rheology_2018} 
or three~\citep{nagy_rheology_2017,nath_rheology_2018} dimensions of space.

In Fig.~\ref{fig:jamming}, I also follow the same procedure for determining the jamming volume fraction 
for the frictional branch $\phi_\mathrm{J}^1$, 
by fitting the viscosity maximum at large stresses to 
$\eta_\mathrm{max} \sim (\phi_\mathrm{J}^1 - \phi)^{-2} \equiv \Delta \phi_1^{-2}$.
I here find that the maximum observed for $\phi_\mathrm{J}^1$ for $\mu_{\mathrm{p}} = 0.5$ 
is much less pronounced than the one for $\phi_\mathrm{J}^0$, and it even absent for $\mu_{\mathrm{p}} = 1$, 
this time in good agreement with what is measured for dry granular systems~\citep{trulsson_rheology_2018,nath_rheology_2018}.

The evolutions of $\phi_\mathrm{J}^0$ and $\phi_\mathrm{J}^1$ as a function of $\alpha$ are such that 
the contrast $\phi_\mathrm{J}^0 - \phi_\mathrm{J}^1$ is maximal in the range $\alpha = 1.5-2$. 
The larger this contrast is, the wider is the range of volume fractions over which the system exhibits 
discontinuous shear thickening and shear jamming.
It is intriguing that cornstarch particles, 
from which it is fairly easy to obtain a shear thickening suspension without finely tuning the volume fraction, 
seem to have aspect ratios in this range~\citep{Brown_2009,Fall_2012}.

\subsection{Normal stress differences}

\begin{figure}[t]
  \centering
  \includegraphics[width=0.9\textwidth]{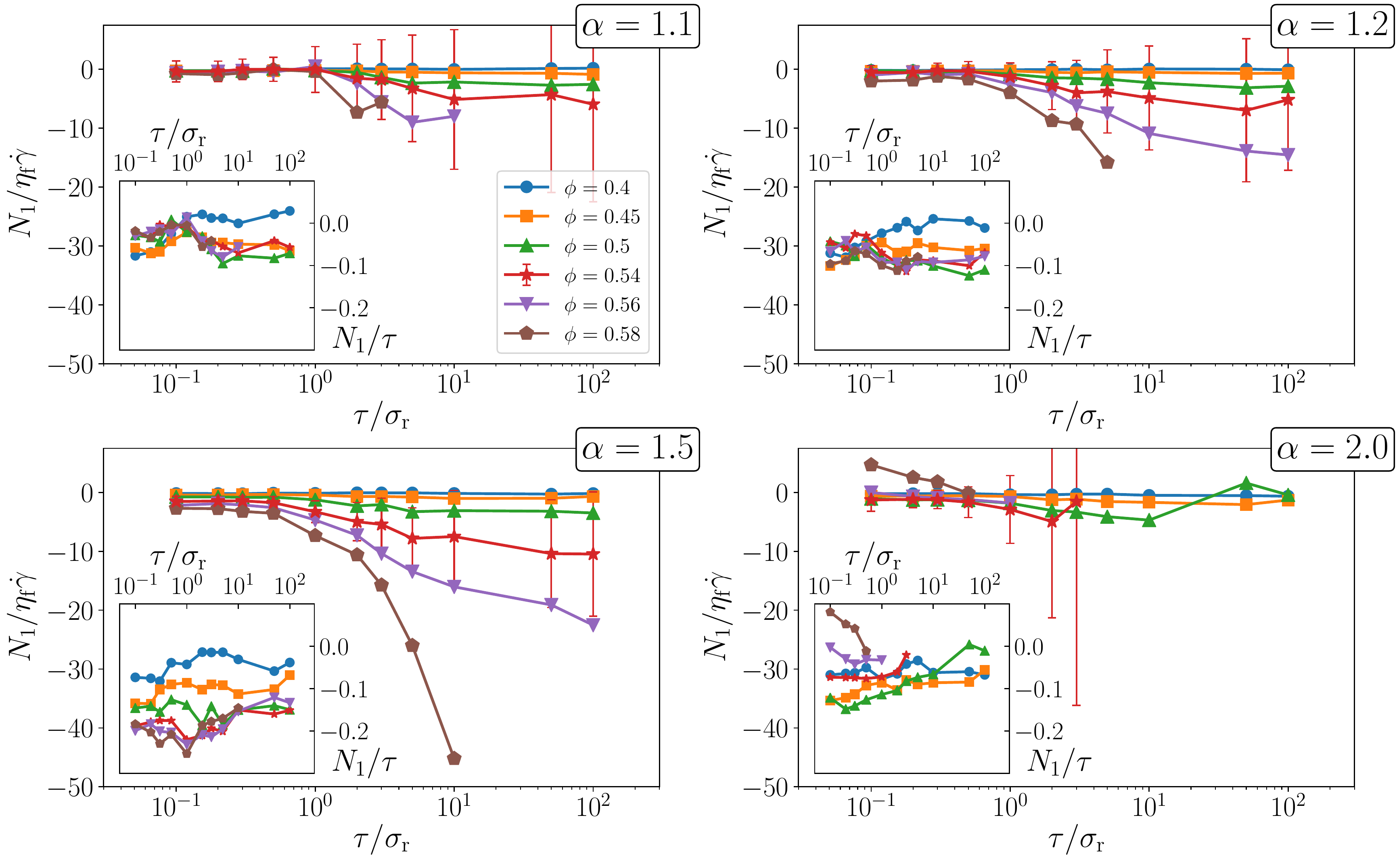}
  \caption{First normal stress difference viscosity $N_1/(\eta_\mathrm{f}\dot\gamma)$  as a function of the applied stress 
  $\tau/\sigma_\mathrm{r}$ for dimers with aspect ratios $\alpha=1.1$ (top left), 
  $\alpha=1.2$ (top right), $\alpha=1.5$ (bottom left), and $\alpha=2$ (bottom right). For each aspect ratio, I show several volume fractions $\phi=0.4,\ 0.45,\ 0.5,\ 0.54,\ 0.56$ and $0.58$. For the $\phi=0.54$ data, error bars give the standard deviations of the time series in steady state. These are representative of the standard deviations at other volume fractions, not shown here to preserve readability. In insets, the same data, but plotted as $N_1/\tau$ as a function of $\tau/\sigma_\mathrm{r}$.}\label{fig:N1}
\end{figure}
I now turn to the normal stress differences $N_1 = \Sigma_{11} - \Sigma_{22}$ and $N_2 = \Sigma_{22} - \Sigma_{33}$. 
In Fig.~\ref{fig:N1}, I show the first normal stress difference viscosity $N_1/(\eta_\mathrm{f}\dot\gamma)$ as a function of the applied stress $\tau/\sigma_\mathrm{r}$.
I find it negative (or small and positive) at all stresses for the smaller values of the aspect ratio, $\alpha=1.1, 1.2$ and $1.5$, 
in contrast with simulations performed on suspensions of spherical particles~\citep{Mari_2014,seto_normal_2018,gallier_rheology_2014}.
(I do not reproduce data for spherical particles in Fig.~\ref{fig:N1} to preserve the readability of the figure.)
Except for the smallest aspect ratio studied ($\alpha=1.1$),
I find no sign of an upturn to eventual positive $N_1$ at large stresses for large volume fractions, 
even for the most abruptly thickening cases.
(It is known that for the present simulation method, the value of $N_1$ is sensitive to the values 
of the spring stiffnesses used in the contact model, with stiffer springs leading to a smaller $N_1$~\citep{seto_normal_2018}. 
The stiffness values used in the present work are around twice larger than what were used in~\citet{Mari_2014}. 
This implies a difference of order $0.01$ on the measured values of $N_1/\tau$~\citep{seto_normal_2018}, 
which cannot account for the difference between spherical and non-spherical particles discussed here.)
For $\alpha=2$ however, positive values of $N_1$ are observed in the shear thickened state for $\phi=0.5$, 
and an upturn is also visible for $\phi=0.54$. 
Unfortunately, for $\phi=0.54$ I could not simulate at larger stresses for large enough strains to report reliable values for $N_1$.
However, even for $\alpha=2$ in this case the values of $N_1/\tau$ have a smaller amplitude than what can be observed for spheres.

This difference is confirmed by the behavior of $N_1/\tau$, in the insets of Fig.~\ref{fig:N1}.
This quantity is clearly showing an increasing trend as a function of $\tau/\sigma_\mathrm{r}$ 
for spherical particles~\citep{Mari_2014}, 
whereas for the suspensions of dimers, even though data have a significant scatter $N_1/\tau$ 
appears rather insensitive to the value of $\tau/\sigma_\mathrm{r}$.
Moreover, $N_1/\tau$ is smaller in amplitude for dimers with $\alpha=1.1, 1.2$ and $2$ than for spheres~\citep{Mari_2014}.
Only dimers with $\alpha=1.5$ stand out, with $N_1/\tau$ values of significantly larger amplitudes.
Finally, note that for $\alpha=2$ I can also observe positive $N_1$ values in the shear thinning regime at the lowest stresses, 
for $\phi=0.56$ and $\phi=0.58$.

\begin{figure}[t]
  \centering
  \includegraphics[width=0.9\textwidth]{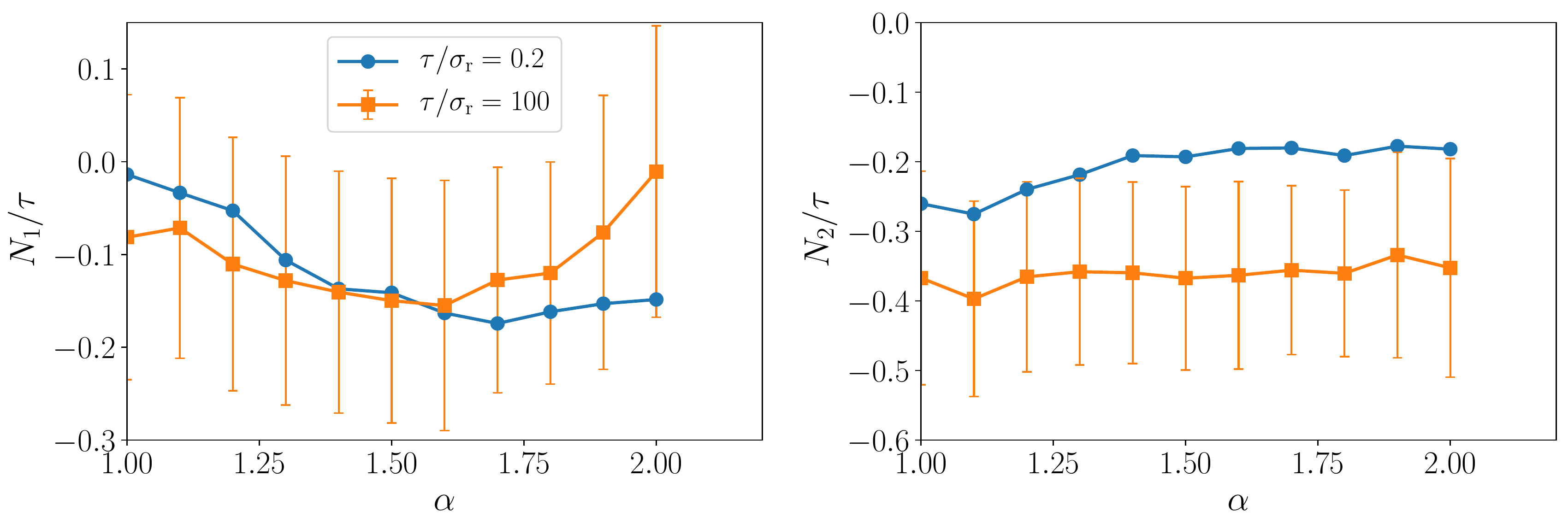}
  \caption{First (left panel) and second (right panel) normal stress differences normalized by the applied stress $N_1/\tau$ and $N_2/\tau$ as a function of the aspect ratio $\alpha$ for a volume fraction $\phi=0.5$ and two stress values $\tau/\sigma_\mathrm{r} = 0.2$ and $\tau/\sigma_\mathrm{r} = 100$.
  For readability, I only show as error bars the standard deviations on the $\tau/\sigma_\mathrm{r} = 100$ data. The standard deviation on the $\tau/\sigma_\mathrm{r} = 0.2$ data
  are of similar magnitude.}\label{fig:N12_alpha}
\end{figure}

A negative $N_1$ at small $\alpha$ contrasts with what is found experimentally and numerically 
for non-Brownian suspensions of large fibers with $\alpha\gg 1$~\citep{snook_normal_2014,bounoua_normal_2016}.
Suspensions of large $\alpha$ particles show a positive $N_1$, which amplitude can be twice as large as the one of $N_2$.
The data suggest that the boundary between these the negative $N_1$ and positive $N_1$ regimes is around $\alpha=2$.
Indeed, if I restrict myself to $\phi=0.5$ for two values of the applied stress, $\tau/\sigma_\mathrm{r} = 0.2$ 
(close to the viscosity minimum in the frictionles state) and $\tau/\sigma_\mathrm{r} = 100$ (frictional state), 
I see that in both cases $N_1/\tau$ shows a negative minimum as a function of $\alpha$ around $\alpha=1.6-1.7$, 
as shown in the left panel of Fig.~\ref{fig:N12_alpha}. 
For the frictional state, $N_1/\tau$ is about to turn positive for $\alpha=2$, 
while this might also happen for larger $\alpha$ values in the frictionless state.

\begin{figure}[t]
  \centering
  \includegraphics[width=0.9\textwidth]{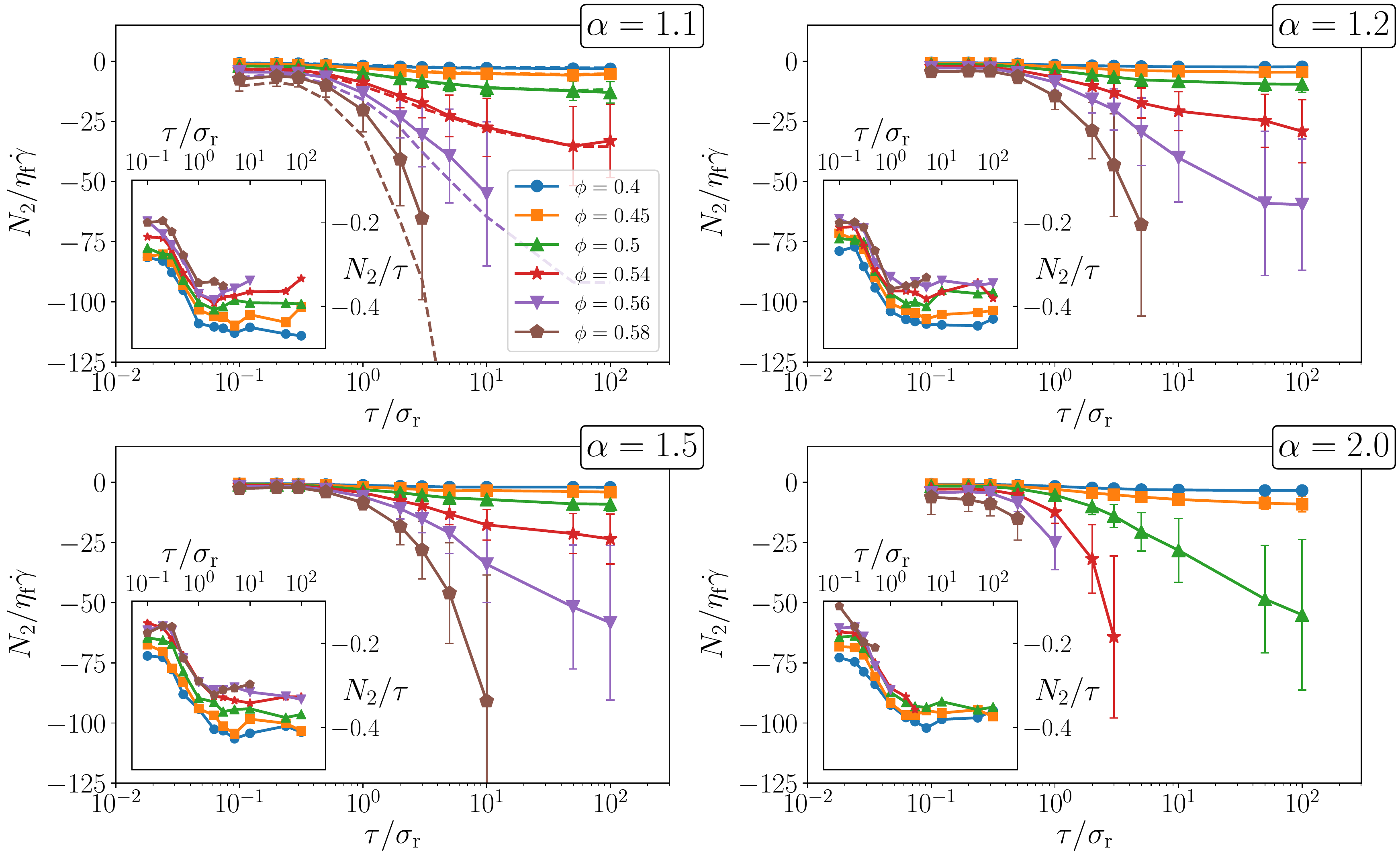}
  \caption{Second normal stress difference viscosity $N_2/(\eta_\mathrm{f}\dot\gamma)$ as a function of the applied stress 
  $\tau/\sigma_\mathrm{r}$ for spheres (aspect ratio $\alpha=1$, top left, dashed lines), and for dimers with aspect ratios $\alpha=1.1$ (top left, symbols)
  $\alpha=1.2$ (top right), $\alpha=1.5$ (bottom left), and $\alpha=2$ (bottom right). 
  For each aspect ratio, I show several volume fractions $\phi=0.4,\ 0.45,\ 0.5,\ 0.54,\ 0.56$ and $0.58$. 
  Error bars give the standard deviations of the time series in steady state.
  In insets, the same data, but plotted as $N_2/\tau$ as a function of $\tau/\sigma_\mathrm{r}$.}\label{fig:N2}
\end{figure}
Regarding $N_2$, in Fig.~\ref{fig:N2}, I show $N_2/(\eta_\mathrm{f}\dot\gamma)$ and $N_2/\tau$ 
as a function of the applied stress $\tau/\sigma_\mathrm{r}$.
As usual for dense non-Brownian suspensions, 
it is negative~\citep{denn_rheology_2014}, and up to 5 times larger than $N_1$ in amplitude.
I find that $N_2/\tau$ decreases when shear thickening occurs, from values of $N_2/\tau \approx -0.2$ up to values around 
$N_2/\tau \approx -0.4$.
This is quite surprisingly similar to suspensions of spherical particles~\citep{Mari_2014}, 
despite the orientational ordering on the system which I will show in a later section.
Indeed, I find a rather mild dependence of $N_2/\tau$ on the aspect ratio, 
as shown in the right panel of Fig.~\ref{fig:N12_alpha} for the same conditions than the $N_1/\tau$ data in the left panel.
It appears to slowly decrease in amplitude with $\alpha$, but this trend saturates for $\alpha\gtrsim 1.5$, 
both in the frictionless and frictional states.
Actually, this extends to much larger aspect ratio, as the amplitude of $N_2$ was also found to be a few tenths of the one of $\tau$ 
even for $\alpha\gg 1$~\citep{snook_normal_2014,bounoua_normal_2016}.

There seems to be a systematic trend with the volume fraction, 
with more concentrated systems showing a smaller $N_2/\tau$ in amplitude. 
The reduction can reach around 25\% from $\phi=0.4$ to $\phi=0.58$ in the thickened state.
For suspensions of spherical particles, this trend is opposite (the amplitude of $N_2/\tau$ increases with $\phi$) for moderate $\phi$~\citep{gallier_rheology_2014}, but $N_2/\tau$ saturates close to jamming~\citep{Mari_2014}.
Interestingly, the amplitude of $N_2/\tau$ was also found to increase with $\phi$ for large aspect ratios~\citep{snook_normal_2014,bounoua_normal_2016}, although usually for volume fractions quite far below jamming.

\subsection{Macroscopic friction coefficient}

\begin{figure}[t]
  \centering
  \includegraphics[width=0.9\textwidth]{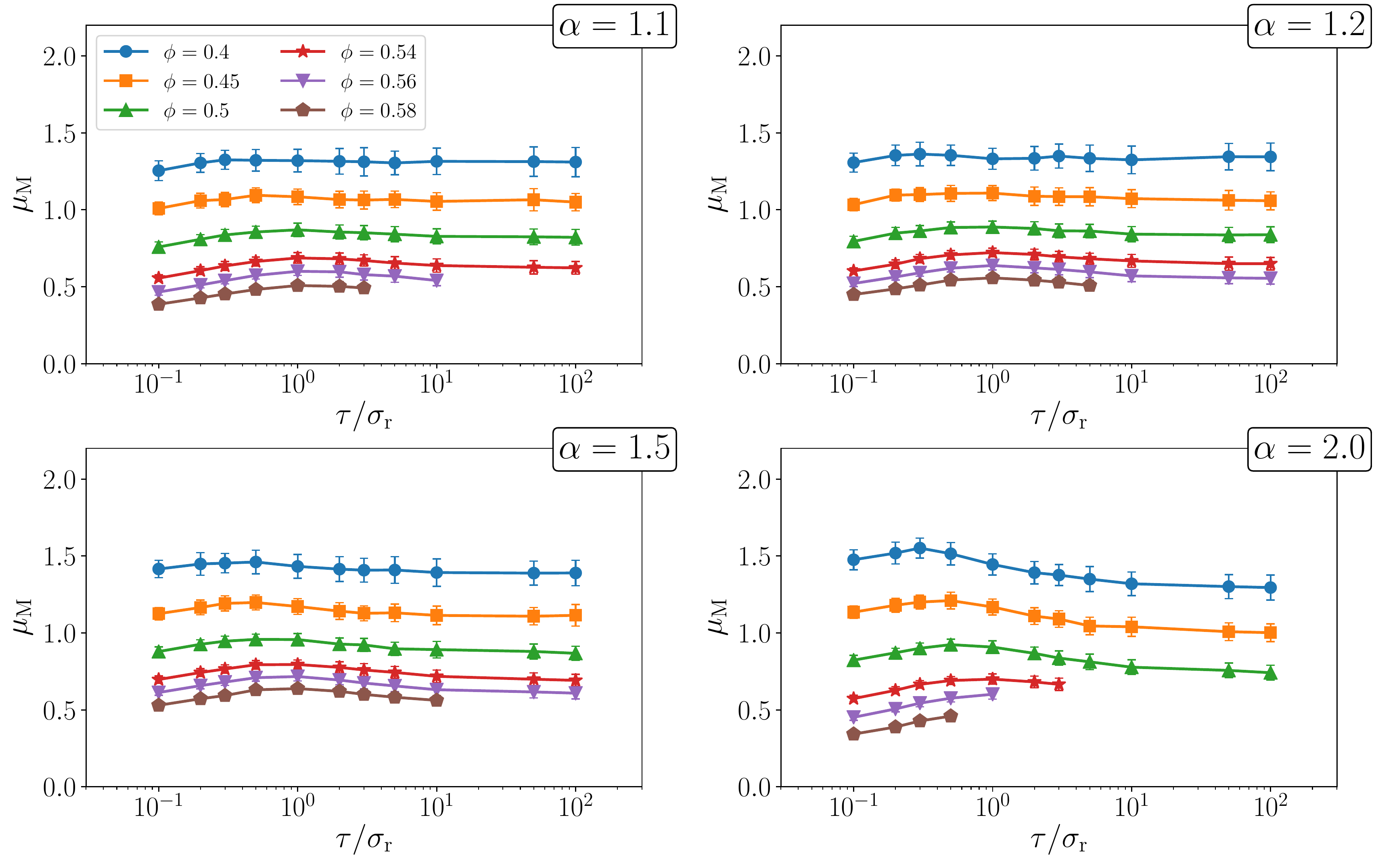}
  \caption{Macroscopic friction coefficient $\mu_\mathrm{M}$ as a function of the applied stress 
  $\tau/\sigma_\mathrm{r}$ for dimers with aspect ratios $\alpha=1.1$ (top left)
  $\alpha=1.2$ (top right), $\alpha=1.5$ (bottom left), and $\alpha=2$ (bottom right). 
  For each aspect ratio, I show several volume fractions $\phi=0.4,\ 0.45,\ 0.5,\ 0.54,\ 0.56$ and $0.58$. 
  Error bars give the standard deviations of the time series in steady state.}\label{fig:mu_stress}
\end{figure}

I now compute the macroscopic friction coefficient 
\begin{equation}
\mu_\mathrm{M} = \tau/P_\mathrm{p},
\end{equation}
where $P_\mathrm{p}$ is the particle pressure, here defined as $P_\mathrm{p} = \mathrm{Tr} \bm{\Sigma}/3$.

In Fig.~\ref{fig:mu_stress}, I show $\mu_\mathrm{M}$ as a function of the applied stress $\tau/\sigma_\mathrm{r}$. 
For all aspect ratios, I find that $\mu_\mathrm{M}$ is much more sensitive to changes in volume fraction 
than to changes in stress. 
The general trend is that $\mu_\mathrm{M}$ decreases with increasing volume fraction, 
as is observed for simpler rate-independent suspensions~\citep{Boyer_2011}.
At fixed $\phi$ and $\alpha$, $\mu_\mathrm{M}$ first increases with $\tau$, 
on a range of low stresses encompassing the shear-thinning regime as well as the start of the shear thickening regime 
as seen in Fig.~\ref{fig:rheology_rate}.
This means that the normal stresses increase slower than the shear stress around the onset of shear thickening.
Finally, for larger applied stresses, $\mu_\mathrm{M}$ decreases again and stabilizes to a plateau value 
in the frictional branch. 
This non-monotonic behavior of $\mu_\mathrm{M}(\tau)$ is more prominent at larger $\phi$ and larger $\alpha$ values.

\begin{figure}[t]
  \centering
  \includegraphics[width=0.9\textwidth]{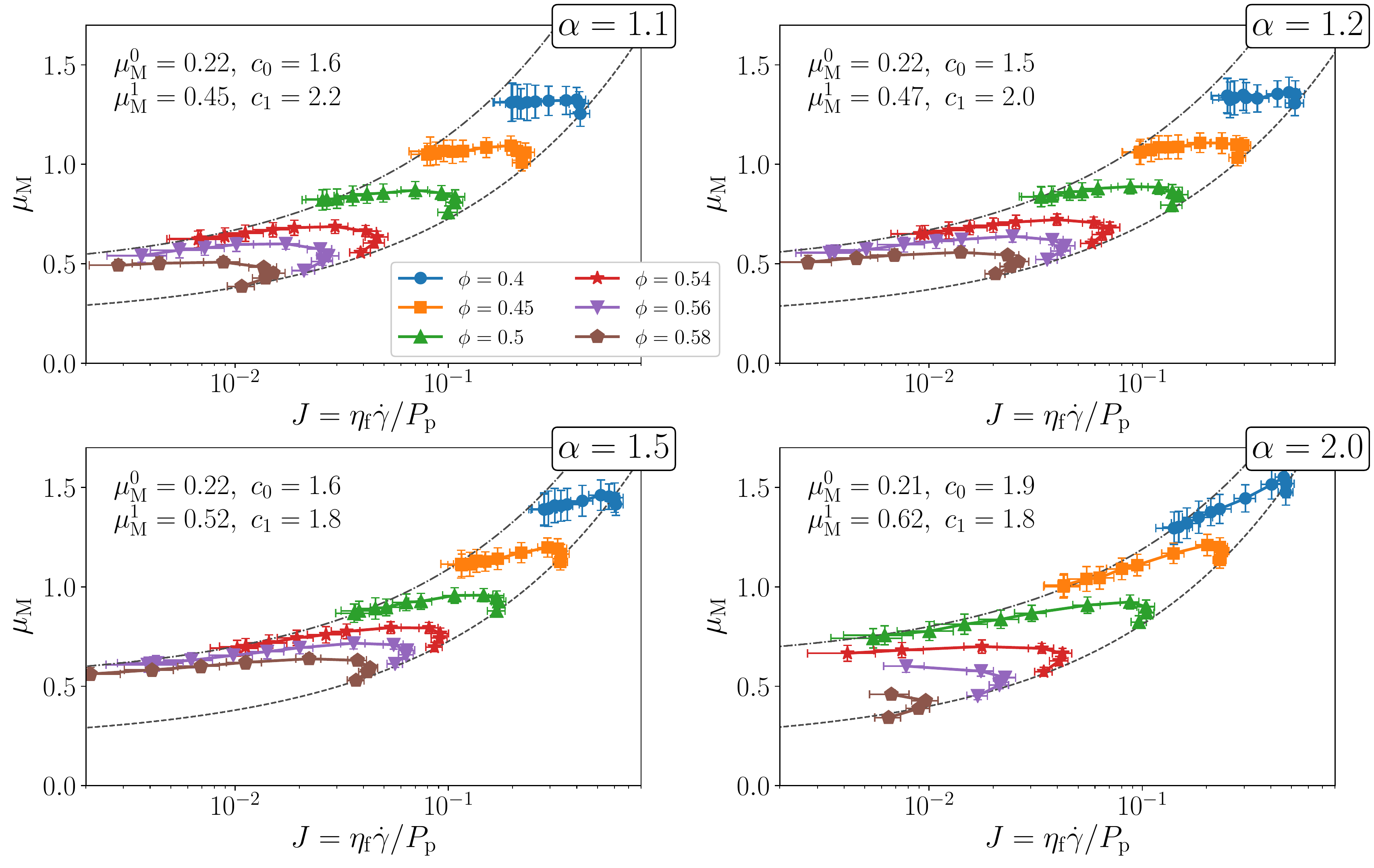}
  \caption{Macroscopic friction coefficient $\mu_\mathrm{M}$ as a function of the viscous number 
  $J = \frac{\eta_\mathrm{f}\dot\gamma}{P_\mathrm{p}}$ for aspect ratios $\alpha=1.1$ (top left), 
  $\alpha=1.2$ (top right), $\alpha=1.5$ (bottom left), and $\alpha=2$ (bottom right). 
  For each aspect ratio, I show several volume fractions $\phi=0.4,\ 0.45,\ 0.5,\ 0.54,\ 0.56$ and $0.58$. 
  Error bars give the standard deviations in $\mu_\mathrm{M}$ and $J$ of the time series in steady state.
   In black dashed (respectively dotted-dashed) lines, fits of the small stress data to Eq.~\ref{eq:mu_m_0} (resp.~Eq.~\ref{eq:mu_m_1}), with the fit parameters indicated in the top left of each figure.}\label{fig:mu_J}
\end{figure}

It is also possible to look at the same data as a function of the viscous number $J = \frac{\eta_\mathrm{f}\dot\gamma}{P_\mathrm{p}}$~\citep{Boyer_2011}.
For rate-independent suspensions, $\mu_\mathrm{M}$ is a function of $J$~\citep{Boyer_2011,Trulsson_2012,gallier_rheology_2014,ness_flow_2015,amarsid_viscoinertial_2017,seto_normal_2018,chevremont_quantitative_2019}, 
which monotonicity has been recently argued to depend on the interparticle friction coefficient~\citep{perrin_interparticle_2019}.
Also, for these suspensions, one can measure $\mu_\mathrm{M}(J)$ in two equivalent ways: 
either in a usual, fixed volume, rate- (or stress-) controlled rheometer, by measuring independently $P_\mathrm{p}$ (hence $J$)  and $\mu_\mathrm{M}$ for several $\phi$, or in a fixed pressure, rate-controlled rheometer 
by measuring $\mu_\mathrm{M}$ and $\phi$ for several $J$~\citep{Boyer_2011}.

For shear thickening suspensions, the rheology depends on an extra dimensionless number $\tau/\sigma_{\mathrm{r}}$, that is,
$\mu_\mathrm{M}$ is a function of both $J$ and $\tau/\sigma_{\mathrm{r}}$.
When performing rheometry at fixed volume and varying the applied stress, one does not follow the same path in 
the $J, \tau/\sigma_{\mathrm{r}}$ parameter space then when performing rheometry at fixed pressure and varying the applied rate. 
Hence, when results for $\mu_\mathrm{M}$ are reported as a function of $J$ only, they need not fall on the same curve in the two cases.
In earlier simulations of shear thickening suspensions under controlled pressure, 
\citet{dong_analog_2017} (for the Critical Load Model~\citep{Mari_2014}) 
and~\citet{kawasaki_discontinuous_2018} (for Brownian frictional hard spheres) 
observed $\mu_\mathrm{M}$ as a function of $J$ (sometimes non-monotonic), 
whether $J$ is controlled by fixing the pressure and varying the rate or fixing the rate and varying the pressure. 
In the case of fixed volume and imposed shear stress, 
I find that $\mu_\mathrm{M}$ is not a function of $J$ (and neither is $J$ a function of $\mu$), 
and this for all aspect ratios and volume fractions I explored,
as shown in Fig.~\ref{fig:mu_J}.
Instead, I find that the relation between $\mu_\mathrm{M}$ and $J$ has a crescent shape. 
This difference highlights the importance of considering the full dependence 
of $\mu_\mathrm{M}$ on both $J$ and $\tau/\sigma_{\mathrm{r}}$.

Nonetheless, in the frictionless branch (or more, precisely, at applied stresses around 
$\tau/\sigma_\mathrm{r} = 0.2$ where the viscosity minimum is observed), 
and in the frictional branch (at large $\tau/\sigma_\mathrm{r}$), one can expect that rate-independence
$\mu_\mathrm{M}(J)$ takes values close to a rate-independent frictionless (respectively frictional) system.
I therefore show in Fig.~\ref{fig:mu_J} fits to the small $J$ asymptotical form of $\mu_\mathrm{M}(J)$ 
proposed by~\citet{Boyer_2011} for the low stress data 
\begin{equation}
\mu_\mathrm{M}^0(J) = \mu_\mathrm{M, c}^0 + c_0 J^{1/2},
\label{eq:mu_m_0}
\end{equation}
and for the large stress data
\begin{equation}
\mu_\mathrm{M}^1(J) = \mu_\mathrm{M, c}^1 + c_1 J^{1/2},
\label{eq:mu_m_1}
\end{equation}
where I adjust $\mu_\mathrm{M, c}^0$ and $c_0$ (respectively $\mu_\mathrm{M, c}^1$ and $c_1$).
I find that the frictionless rheology $\mu_\mathrm{M}^0(J)$ is much less dependent on the aspect ratio 
than the frictional rheology $\mu_\mathrm{M}^1(J)$.
The most spectacular effect of an increase in aspect ratio is the large increase in onset friction coefficient 
$\mu_\mathrm{M, c}^1$, from $\mu_\mathrm{M, c}^1 = 0.45$ for $\alpha=1.1$ to $\mu_\mathrm{M, c}^1 = 0.62$ 
for $\alpha=2$.

\subsection{Orientational order}

\begin{figure}[t]
  \centering
  \includegraphics[width=0.9\textwidth]{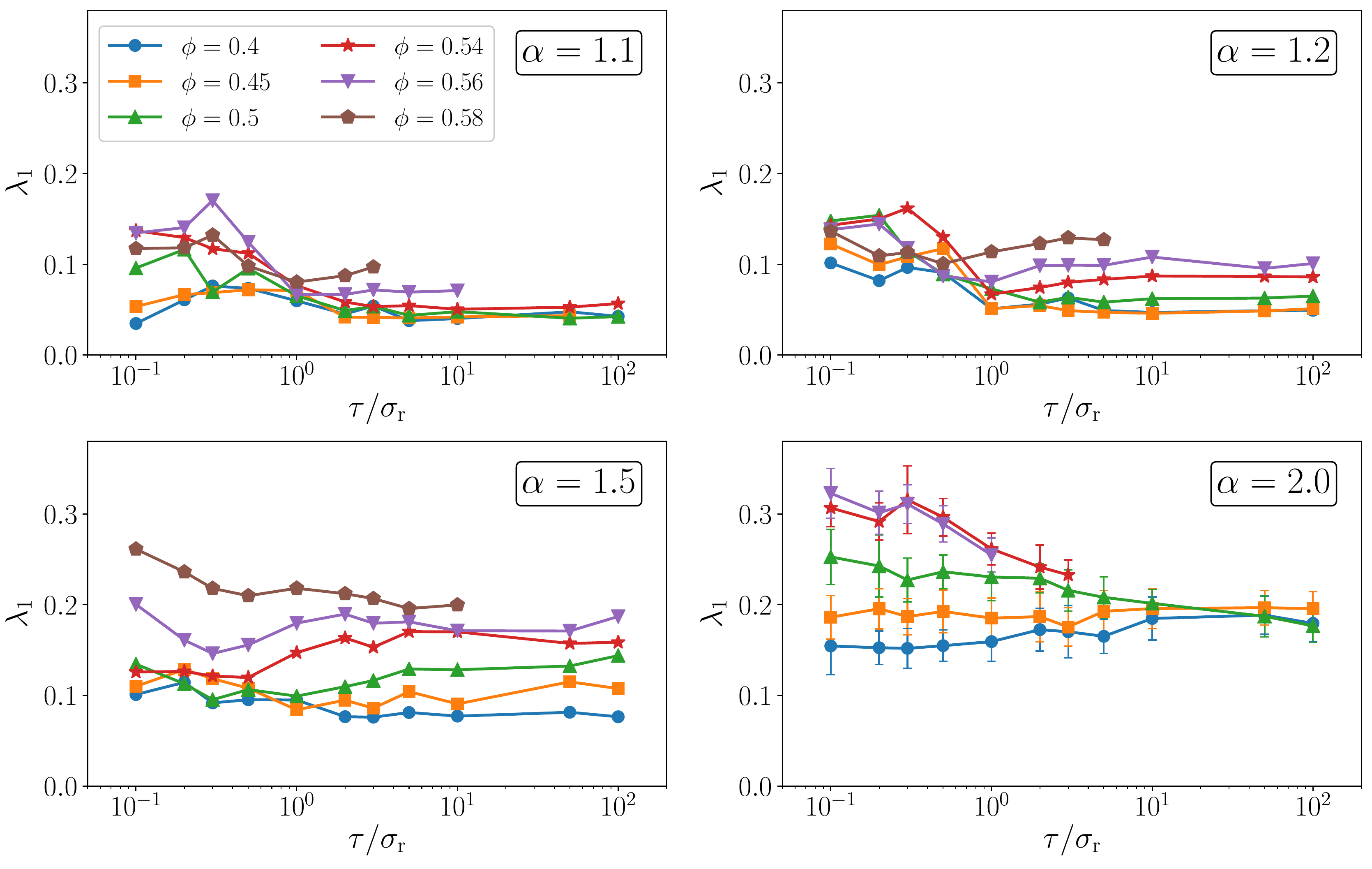}
  \caption{Nematic order parameter $\lambda_1$ (see text for definition) as a function of the applied stress 
  $\tau/\sigma_\mathrm{r}$, for aspect ratios $\alpha=1.1$ (top left), 
  $\alpha=1.2$ (top right), $\alpha=1.5$ (bottom left), and $\alpha=2$ (bottom right). 
  For each aspect ratio, I show several volume fractions $\phi=0.4,\ 0.45,\ 0.5,\ 0.54,\ 0.56$ and $0.58$. 
  I show the standard deviation of the time series with error bars on the $\alpha=2$ data. Standard deviations for other aspect ratios are of similar magnitude, but were omitted for readability.}\label{fig:order_parameter}
\end{figure}

To characterize the average orientation of dimers, which is a nematic quantity, I use the fabric tensor
\begin{equation}
  \bm{Q} = \frac{3}{2N_D} \sum_\alpha \bm{n}_\alpha \bm{n}_\alpha - \bm{I}/3,
\end{equation}
where the sum runs over the dimers, and $\bm{n}_\alpha$ is the unit vector along the center-to-center vector between the two spheres making the dimer. 
One can perform an eigenvector decomposition of this tensor as
\begin{equation}
  \bm{Q} = \lambda_1 \bm{u}_1\bm{u}_1 + \lambda_2 \bm{u}_2\bm{u}_2 + \lambda_3 \bm{u}_3\bm{u}_3,
\end{equation}
where $\bm{u}_{1,2,3}$ are the eigenvectors and $\lambda_1>\lambda_2>\lambda_3$ the associated eigenvalues of $\bm{Q}$.
I use the largest eigenvalue, $\lambda_1$, as a scalar order parameter: a completely isotropic state has $\lambda_1 = 0$, 
and a state where all dimers have the same orientation has $\lambda_1 = 1$.
In simple shear, from symmetry considerations one expects the eigenvectors, and in particular the so-called director $\bm{u}_1$, 
to lie either in the shear plane, or along the vorticity direction.
To represent the director $\bm{u}_1$, I will follow earlier literature and use its spherical coordinates with the vorticity as the zenith direction, 
with $\varphi$ the angle between the director and the vorticity direction and $\theta$ the angle between the flow direction 
and the projection of the director on the flow plane~\citep{campbell_elastic_2011,guo_numerical_2012,nagy_rheology_2017,nath_rheology_2018,marschall_orientational_2019}.
Because the order is nematic, I use directors such that $0<\theta<\pi$. 
Finally, because of the ambiguous definition of $\theta$ when $\varphi=0$, there is a lot of noise on the data when $\varphi$ is small. 
I chose to report the values of $\theta$ only for $\varphi>0.1$, when I am confident that the average $\theta$ value has converged.

\begin{figure}[t]
  \centering
  \includegraphics[width=0.9\textwidth]{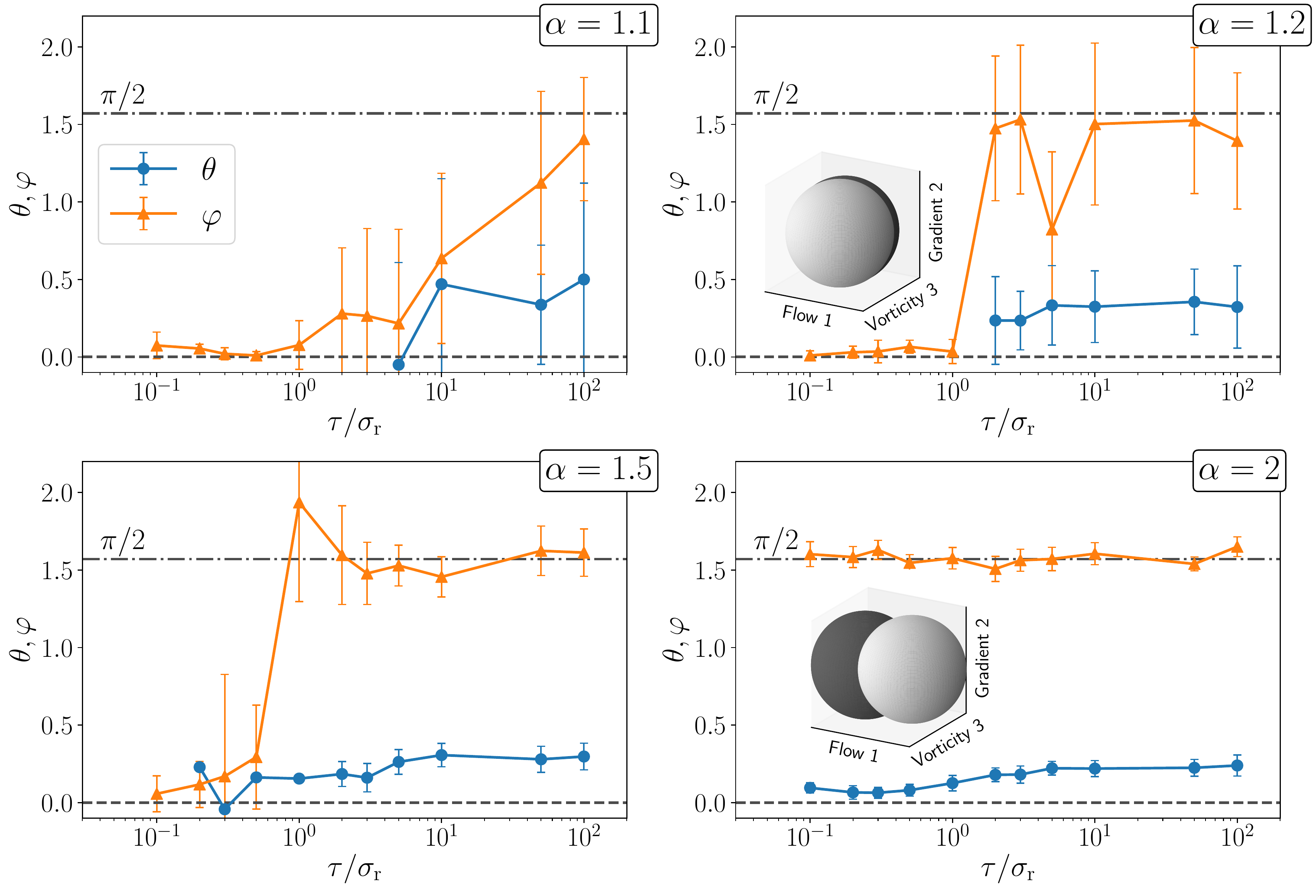}
  \caption{Angles $\theta$ and $\varphi$ characterizing the director (see text for definitions) as a function of the applied stress 
  $\tau/\sigma_\mathrm{r}$, for aspect ratios $\alpha=1.1$ (top left), 
  $\alpha=1.2$ (top right), $\alpha=1.5$ (bottom left), and $\alpha=2$ (bottom right), at $\phi=0.5$. 
  Error bars represent the standard deviations on the time series of the angles.
  I also show a dimer with the average orientation taken at the lowest simulated stress for $\alpha=1.2$, and at the highest simulated stress for $\alpha=2$ , 
  illustrating the orientational transition from vorticity to flow alignment.}\label{fig:director}
\end{figure}

\begin{figure}[t]
  \centering
  \includegraphics[width=0.9\textwidth]{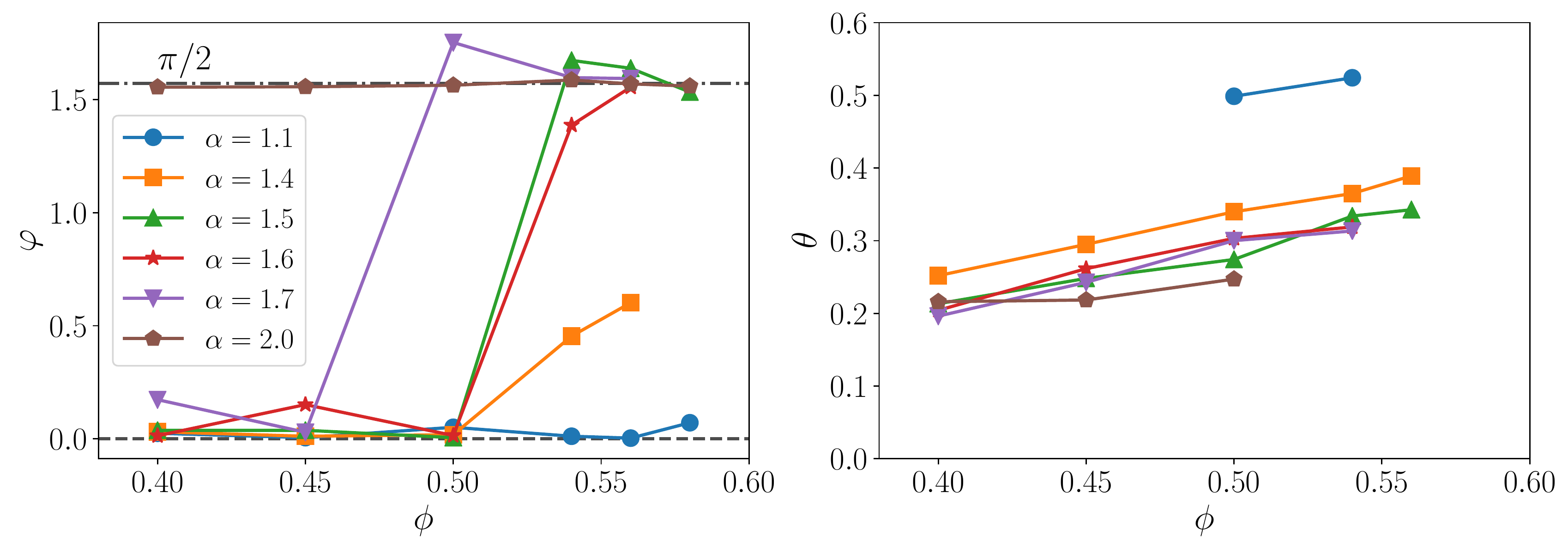}
  \caption{\textbf{Left:} Angle $\varphi$ as a function of the volume fraction $\phi$ for applied stress 
  $\tau/\sigma_\mathrm{r} = 0.2$ (frictionless state) and several aspect ratios.
  \textbf{Right:} Angle $\theta$ as a function of the volume fraction $\phi$ for applied stress 
  $\tau/\sigma_\mathrm{r} = 100$ (frictional state) and the same aspect ratios.}\label{fig:angles_phi}
\end{figure}

I first show the order parameter $\lambda_1$ in Fig.~\ref{fig:order_parameter} as a function of the applied stress, 
for several values of $\alpha$ and $\phi$.
The suspension is more ordered for larger aspect ratios and larger volume fractions, 
as in dry systems~\citep{reddy_orientational_2009,guo_granular_2013,farhadi_dynamics_2014,nagy_rheology_2017,trulsson_rheology_2018}.
For a given $\alpha$ and $\phi$, the system is usually more ordered below than above shear thickening, except perhaps for 
$\alpha = 1.5$, where the order is rather insensitive to shear thickening.
This trend is consistent with earlier observations in dry granular systems, 
for which frictionless particles order more than frictional ones~\citep{trulsson_rheology_2018,nath_rheology_2018}.
The values globally are lower than for dry systems~\citep{nagy_rheology_2017,trulsson_rheology_2018,nath_rheology_2018}, 
both in the frictionless and the frictional regimes, which could stem from the presence of lubrication, which generates 
extra torques on the particles, tending to disturb alignment.

I show in Fig.~\ref{fig:director} the director associated with this order.
At large aspect ratios and large stresses, the director aligns in the shear plane ($\varphi\approx\pi/2$), 
with a positive angle $\theta$ with respect to the flow direction.
It is consistent with SANS data obtained on a shear-thickening suspension of particles 
with $\alpha\approx 7$ by~\citet{egres_rheology_2005}, and confocal microscopy observations 
on a shear-thickening suspension of rods with $\alpha\approx 10$ 
by~\citet{rathee_unraveling_2019}. 
This angle has a parameter dependence with the same trends than for systems of dry elongated particles.
It increases when shear thickening occurs, i.e.~when friction increases 
(for dry systems, see~\citep{trulsson_rheology_2018}).
It also increases with the volume fraction, as shown on the right panel of Fig.~\ref{fig:angles_phi} 
in the thickened state (for dry systems, see~\citep{farhadi_dynamics_2014,trulsson_rheology_2018}, 
although the opposite trend has also been observed~\citep{reddy_orientational_2009,guo_numerical_2012})
Finally, it decreases with increasing aspect ratio (for dry systems, see~\citep{borzsonyi_orientational_2012,borzsonyi_shear_induced_2012,guo_numerical_2012,guo_granular_2013,nagy_rheology_2017,trulsson_rheology_2018}).

However, for smaller aspect ratios $\alpha=1.1, 1.2$ and $1.5$, at stresses in the shear thinning regime, 
I observe another orientation along the vorticity direction ($\varphi\approx 0$), 
which has never been reported before.
In this situation, the dimers roll around their symmetry axis with the flow. 
Note that this state is usually more ordered than the flow aligned one, as seen in Fig.~\ref{fig:order_parameter}.
I observe this behavior up to an aspect ratio dependent volume fraction, as can be seen in the left panel 
of Fig.~\ref{fig:angles_phi}, where I plot $\varphi$ as a function of $\phi$ for several aspect ratios, in the 
unthickened state with $\tau/\sigma_\mathrm{r} = 0.2$.
The volume fraction below which dimers are vorticity aligned decreases with $\alpha$.
For $\alpha>1.7$, I never observe this alignment for the volume fractions studied here, whereas for $\alpha<1.4$, 
I never get alignment along the flow direction.

\begin{figure}[t]
  \centering
  \includegraphics[width=0.9\textwidth]{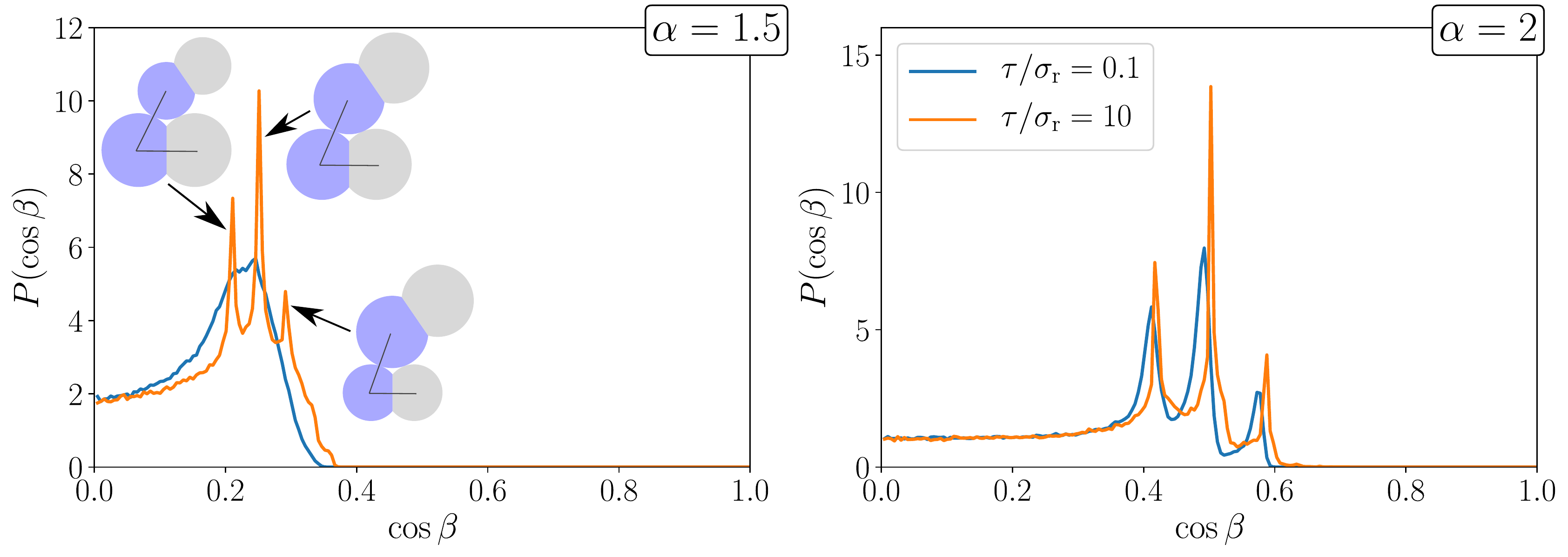}
  \caption{Probability distribution of $\cos \beta$  (see main text for the definition of the angle $\beta$) for stresses 
  $\tau/\sigma_\mathrm{r} = 0.1$ (below shear thickening) and $\tau/\sigma_\mathrm{r} = 10$ (above shear thickening), volume fraction $\phi=0.5$, 
  and aspect ratios $\alpha=1.5$ (left) and  $\alpha=2$ (right).}\label{fig:lock_in}
\end{figure}

The origin of the vorticity alignment does not seem to be tied with the non-convexity of the dimers. 
A typical effect of the non-convexity of the particles is that they can interlock, a situation where a sphere 
of a dimer preferentially stands in the concave part in between the two spheres of a neighboring dimer. 
One can quantify this effect by systematically measuring the angle $\beta$ between the director of a dimer involving spheres $i$ and $j$ 
and the separation vector between one of the dimer's sphere (say $i$) and another sphere $k\neq j$ belonging to an other dimer, that is, 
$\cos \beta = (\bm{r}_j - \bm{r}_i)\cdot (\bm{r}_k - \bm{r}_i) / |\bm{r}_j - \bm{r}_i||\bm{r}_k - \bm{r}_i|$.
I do this for all neighboring spheres with gaps $h^{(i,k)} < 0.1$. 
In Fig.~\ref{fig:lock_in} I show the probability distribution of $\cos\beta$, for $\alpha = 1.5$ and $\alpha = 2$ at a low stress $\tau/\sigma_\mathrm{r} = 0.1$ 
corresponding to vorticity alignment for $\alpha = 1.5$ and flow alignment for $\alpha = 2$, and at a high stress  $\tau/\sigma_\mathrm{r} = 10$ 
corresponding to flow alignment for both aspect ratio.
This distribution would be flat ($P(\cos\beta)$ independent of $\cos\beta$) if the neighboring spheres were uniformly distributed around a dimer.
Instead, at large $\cos\beta$, one sees that there is an excluded solid angle with $P(\cos\beta)\simeq 0$, 
where the partner sphere prevents another particle to align with the dimer director.
The extension of this excluded region of course depends on the aspect ratio; it shrinks with increasing $\alpha$.
More interestingly, out of this excluded region, the distribution systematically has three peaks when the dimers are predominantly flow aligned.
These peaks are located at values of $\cos \beta$ corresponding respectively to the three possible interlocking configurations: 
a large sphere in the concave part of a small dimer (peak at the smallest $\cos \beta$), a large sphere in a large dimer or a small sphere in a small dimer (middle peak), 
and a small sphere in a large dimer (rightmost peak, largest $\cos\beta$ value).
Now, in all the cases where spheres are vorticity aligned (here $\alpha = 1.5$ and $\tau/\sigma_\mathrm{r} = 0.1$), there is no observable peak, just a broad maximum. 
Said otherwise, there is an anti-correlation between vorticity alignment and interlocking, strongly suggesting that the vorticity alignment 
is not specific to non-convex the particles.

\begin{figure}[t]
  \centering
  \includegraphics[width=0.9\textwidth]{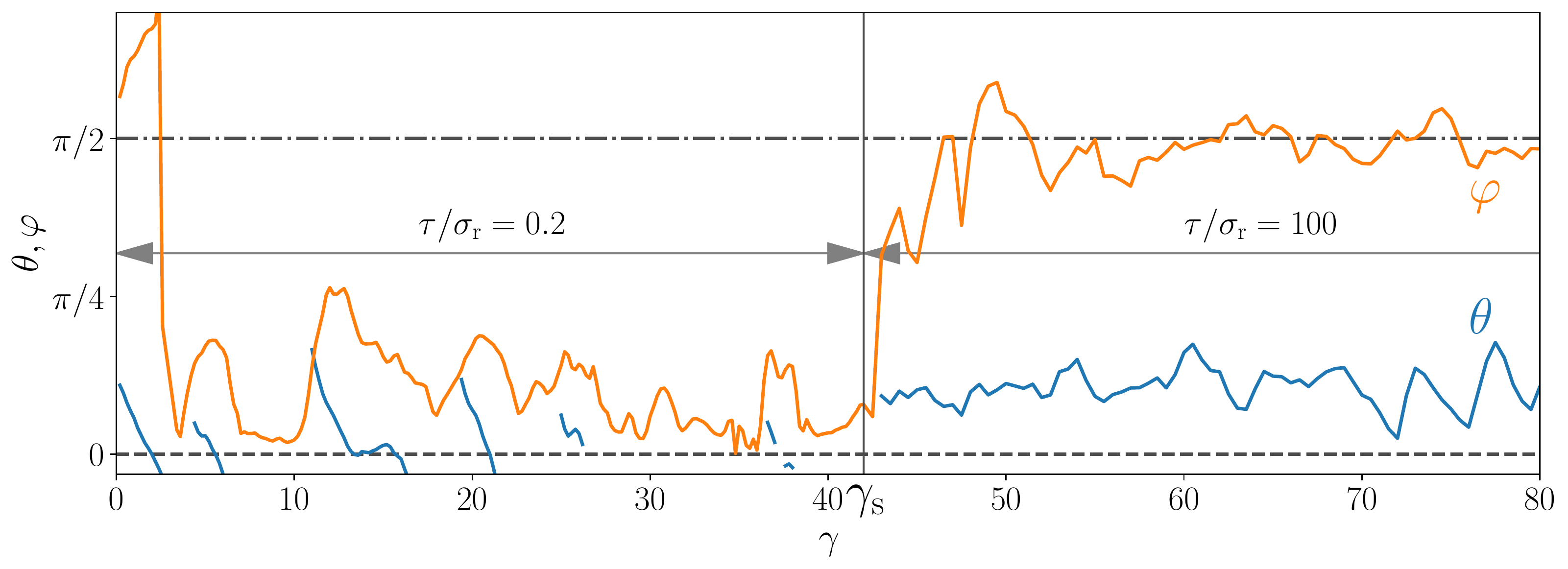}
  \caption{Angles $\varphi$ and $\theta$ as a function of strain, for a simulation where I apply a small stress $\tau/\sigma_\mathrm{r}=0.2$ up to strain $\gamma_\mathrm{s}$, and then a large stress $\tau/\sigma_\mathrm{r}=100$. Here, $\alpha=1.5$ and $\phi=0.5$.}\label{fig:stress_switch}
\end{figure}
Interestingly, the strain needed to acquire a new orientation is quite short. In Fig.~\ref{fig:stress_switch}, I show 
the reorientation dynamics for a suspension with $\alpha=1.5$ and $\phi=0.5$.
Starting up a shear under small stress $\tau/\sigma_\mathrm{r}=0.2$ from an initial configuration slightly biased towards flow alignment, 
the particles orient preferentially towards the vorticity direction during an initial transient of less than five strain units.
At a later strain $\gamma_\mathrm{s}$, the suspension is subject to a stress switch to the frictional state 
at $\tau/\sigma_\mathrm{r}=100$. 
Here again, the director rapidly rotates from the vorticity direction to the shear plane, 
during a transient lasting roughly $5-10$ strain units.

\subsection{Shear thickening by geometric friction}

\begin{figure}[t]
  \centering
  \includegraphics[width=0.9\textwidth]{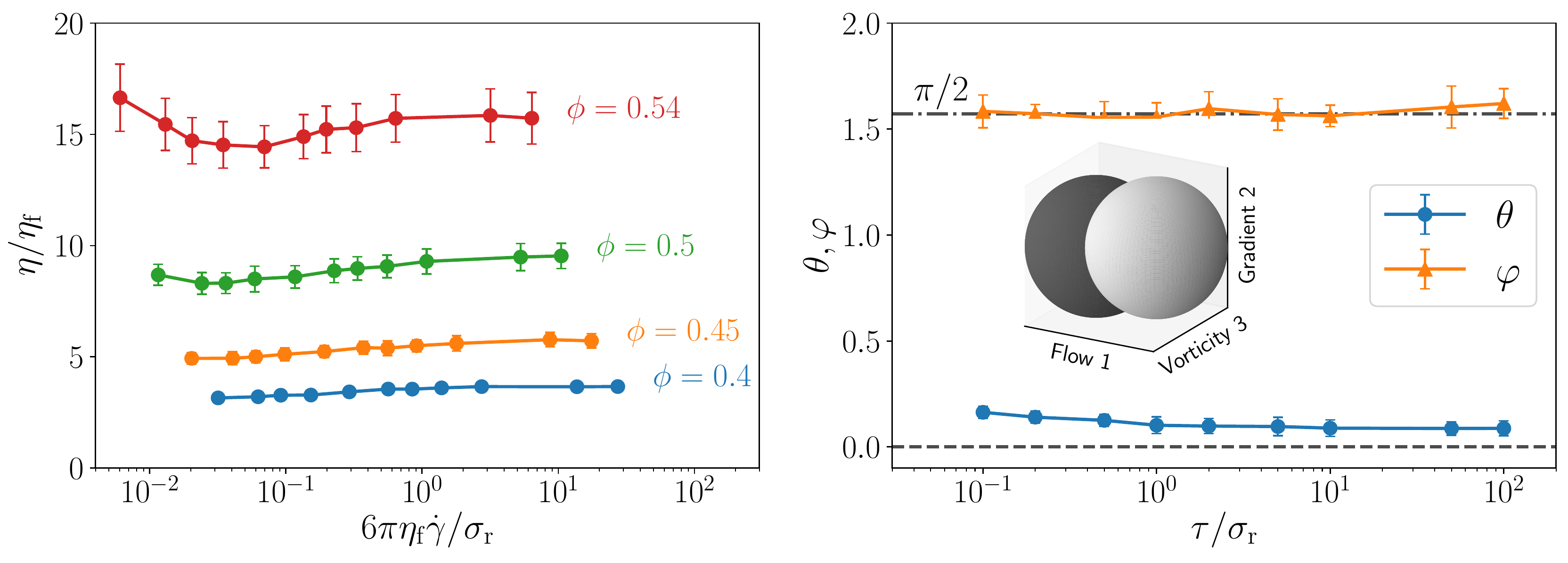}
  \caption{\textbf{Left:} Relative viscosity as a function of the shear rate for frictionless dimers with aspect ratio $\alpha=2$, for several volume fractions $\phi=0.4, 0.45, 0.5$ and $0.54$. \textbf{Right:} Angles $\theta$ and $\varphi$ characterizing the director (see text for definitions) as a function of the applied stress 
  $\tau/\sigma_\mathrm{r}$, for the corresponding simulations at $\phi=0.5$ in the left panel.}\label{fig:frictionless_thickening}
\end{figure}

In this last section, I address the fact that due to their non-convex shape,
under interlocking configurations even frictionless dimers can exchange
forces with a significant component tangential to the dimer orientation, 
akin to a ``geometrical friction''.
(The difference between usual and geometrical friction is probably blurry.
At the scale of the surface roughness, this geometrical locking is possibly 
the cause of actual friction observed at the particle scale; some experiments actually 
probed situations where roughness and particle non-convexity 
are undistiguishable~\citep{hsiao_rheological_2017,hsu_roughness_dependent_2018}.)
Could this geometrical friction be enough to generate shear thickening?

I show in the left panel of Fig.~\ref{fig:frictionless_thickening} the flow curves $\eta(\dot\gamma)$ 
for frictionless ($\mu_\mathrm{p}=0$) dimers with aspect ratio $\alpha=2$. 
I indeed observe a shear thickening, which is more prominent when the volume fraction increases.
I also observe that this thickening is not linked to a significant change in orientational order, as the director 
(shown in the right panel of Fig.~\ref{fig:frictionless_thickening} for $\phi=0.54$) 
is almost constant, staying in the shear plane with a slightly decreasing angle $\theta$ 
across the stress range I probed.
The amplitude of the thickening however remains modest (a viscosity increase of roughly 10\%), 
and is negligeable compared to the effect 
of actual interparticle friction with $\mu_{\mathrm{p}} = 0.5$ as shown in Fig.~\ref{fig:rheology_rate}.

\section{Conclusions}

I reported results of simulations of sheared suspensions of frictional and repulsive hard particles with dimeric shape.
I showed that these suspensions, just like their spherical particle suspensions counterpart, 
undergo a shear thickening which is continuous when the volume fraction is below than a critical value $\phi_\mathrm{c}$, 
and discontinuous above.
The increase of viscosity is for the most part due to the switching of the frictional interactions resulting from the competition 
between applied stress and repulsive forces, although a small proportion is also coming from 
geometrical friction due to the non-convexity of the particles.
Most of the effect of the non-sphericity is quantitative, and captured by the evolution of the usual parameters 
of shear thickening, like the locations of frictionless and frictional jamming points  $\phi_\mathrm{J}^0$ and $\phi_\mathrm{J}^1$, 
and the value of the DST onset $\phi_\mathrm{c}$.
In particular, because $\phi_\mathrm{J}^0$ increases significantly with aspect ratio from $\alpha=1$ (spheres) to $\alpha\approx 1.5$, 
whereas $\phi_\mathrm{J}^1$ barely increases (or even decreases for interparticle friction coefficients $\mu_\mathrm{p} \gtrsim 1$),
at a given $\phi$ the viscosity difference between untickened and thickened states increases with $\alpha$.
The same increase of the viscosity difference happens when one considers spherical particles interacting with rolling friction, 
as opposed to sliding friction only~\citep{mari_force_2019}, which supports the idea that some of the effects of non-sphericity 
can be captured by rolling friction.
Along these quantitative differences between spherical and non-spherical particles, 
a qualitative difference can be detected in the sign of the first normal stress difference, 
which remains negative for aspect ratios $\alpha=1.1, 1.2$, and $1.5$, even when discontinuous shear thickening occurs.

Separately, I studied the ordering of these shear thickening suspensions of dimers, and uncovered its dependence on applied stress. 
Interestingly, the director (i.e.~the principal orientation taken by the particles) is not the same below and above shear thickening 
when the aspect ratio is in between $\alpha=1.4$ and $\alpha=1.7$. 
I indeed found that at small applied stresses and small aspect ratios, 
particles are primarily oriented along the vorticity direction, 
whereas at large stresses and large aspect ratios they are primarily oriented in the shear plane, 
at a finite but small angle $\theta$ with the flow direction.
Only the latter orientation is seen in dry granular systems under simple shear~\citep{campbell_elastic_2011,borzsonyi_orientational_2012,borzsonyi_shear_induced_2012,guo_numerical_2012,guo_granular_2013,nagy_rheology_2017,trulsson_rheology_2018}.

The modeling of this orientational phase diagram is yet to be developed. 
Surely, a theory describing the nematic order parameter should also have a fully tensorial rheology, 
as opposed to the scalar Wyart-Cates theory or (extensions of) $\mu(J)$ rheology. 
For dense supensions, such rheological models are currently
the object of active research, 
but so far attempts were limited to suspensions of spherical particles~\citep{chacko_shear_2018,gillissen_modeling_2018,ozenda_new_2018,singh_constitutive_2018,baumgarten_general_2019,gillissen_constitutive_2019}.

The stress-order coupling could lead to spectacular phenomena.
For rods with large aspect ratios, \citet{rathee_unraveling_2019} recently observed large amplitude viscosity oscillations 
during discontinuous shear thickening concurrent with orientation changes. 
Spikes of large viscosities were associated to an alignment of the rods in the gradient direction,
while low viscosity periods were associated to the rods being along the flow direction.
This behavior is only observed during discontinuous shear thickening under imposed shear stress. 
In this situation, even for spherical particles one observes flow instabilities due to the non-monotonic 
flow curve~\citep{nagahiro_negative_2016,hermes_unsteady_2016,rathee_localized_2017,saint_michel_uncovering_2018,chacko_dynamic_2018}. 
It is probable that elongated particles will considerably enrich this dynamics by coupling nematic order to the stress field.

\appendix

\end{document}